\pdfoutput=1 
\documentclass[preprint2]{aastex}
\usepackage{fancyheadings}
\usepackage[pdftex]{epsfig}
\usepackage{epstopdf}

\newcommand{\tick}{ \hspace{-4pt}$\surd$ \hspace{-4pt}} 
\newcommand{\cross}{ \hspace{-4pt}$\times$ \hspace{-4pt}}

\newcommand{\dg}{$^{\circ}$\,}
\newcommand{\as}{$''$\,}
\newcommand{\am}{$'$\,}
\newcommand{\Kdark}{$K_d$\,}
\newcommand{\etal}{{\it et al.\,}}
\slugcomment{Resubmitted to \pasp.}

\shorttitle{The best site on Earth?}
\shortauthors{Saunders \etal}
\begin{document}

\title{Where is the best site on Earth? \\
Domes A, B, C and F, and Ridges A and B}

\author{Will Saunders$^{1,2}$, Jon S. Lawrence$^{1,2,3}$, John W.V. Storey$^1$, Michael C.B. Ashley$^1$}
\affil{$^1$School of Physics, University of New South Wales\\ $^2$Anglo-Australian Observatory\\ $^3$Macquarie University, New South Wales }
\email{will@aao.gov.au}

\author{Seiji Kato, Patrick Minnis, David M. Winker}
\affil{NASA Langley Research Center}

\author{Guiping Liu}
\affil{Space Sciences Lab, University of California Berkeley}

\author{Craig Kulesa}
\affil{Department of Astronomy and Steward Observatory, University of Arizona}

\begin{abstract}
 
The Antarctic plateau contains the best sites on earth for many forms of astronomy, but none of the existing bases was selected with astronomy as the primary motivation.  In this paper, we try to systematically compare the merits of potential observatory sites. We include South Pole, Domes A, C and F, and also Ridge B (running NE from Dome A), and what we call `Ridge A' (running SW from Dome A). Our analysis combines satellite data, published results and atmospheric models, to compare the boundary layer, weather, aurorae, airglow, precipitable water vapour, thermal sky emission, surface temperature, and the free atmosphere, at each site. We find that all Antarctic sites are likely to be compromised for optical work by airglow and aurorae. Of the sites with existing bases, Dome A is easily the best overall; but we find that Ridge A offers an even better site. We also find that Dome F is a remarkably good site. Dome C is less good as a thermal infrared or terahertz site, but would be able to take advantage of a predicted `OH hole' over Antarctica during Spring.

\end{abstract}


\section{Introduction}
There are now many articles on the characteristics of the various Antarctic sites; for a summary see, for example, Storey (2005) and Burton (2007).  This work attempts to draw together some of these papers, and also unpublished meteorological and other information for these sites, to help characterise what are almost certainly the best sites on Earth for many forms of astronomy. The factors considered in this study are:

\begin{itemize}
\vspace{-9pt}
{\item Boundary layer thickness}
\vspace{-6pt}
{\item Cloud cover}
\vspace{-6pt}
{\item Auroral emission}
\vspace{-6pt}
{\item Airglow}
\vspace{-6pt}
{\item Atmospheric thermal backgrounds }
\vspace{-6pt}
{\item Precipitable water vapour (PWV)}
\vspace{-6pt}
{\item Telescope thermal backgrounds}
\vspace{-6pt}
{\item Free-atmosphere seeing}
\end{itemize}

Of course, different astronomical programs have very different requirements. For high resolution optical work, including interferometry, it is the turbulence characteristics (including seeing, isoplanatic angle and coherence time) that are most important. For wide-field optical work it is seeing, auroral emission, airglow, weather, and sky coverage. For the thermal near-infrared, including \Kdark at 2.4$\mu$m,  it is the thermal backgrounds from sky and telescope. For mid-infrared and terahertz work the PWV is paramount. 

There are other significant issues not covered in this study, for example: sky coverage, daytime use, existing infrastructure, accessibility, telecommunications, and non-astronomical uses.

\section{The possible sites}
The sites where astronomical work has taken place or is under consideration are South Pole, and Domes A, C and F. We have also included Ridge B, running NE from Dome A; according to the digital map of Liu \etal (2001), Ridge B contains a genuine peak at its southern end, which we call Dome B, at (79\dg S, 93\dg E, 3809m). We also consider the ridge leading southwest from Dome A, which we call Ridge A. We do not consider Vostok in this study, as it does not lie on a ridge or dome, and unlike South Pole, does not have extensive available site testing or astronomical data.

\begin{table}[htpb]
\begin{center}
\caption{locations of the sites. Elevation is from Liu \etal (2001). \label{tbl-1}}
\begin{tabular}{lrrr}
\tableline\tableline
 Site & Latitude&Longitude &Elevation\\

\tableline
South Pole &   90\dg S&  0\dg E	&  2800m    \\
Dome A &80.37\dg S  & 77.53\dg E & 4083m  \\		
Dome C &75.06\dg S  &123.23\dg E &  3233m \\
Dome F &77.19\dg S & 39.42\dg E &  3810m  \\
Ridge B	&$\sim$76\dg S &$\sim$94.75\dg E & $\sim$3750m   \\
Dome B	&79.0\dg S &93.6\dg E & 3809m   \\
Ridge A &81.5\dg S  & 73.5\dg E & 4053m  \\		
\tableline
\end{tabular}
\end{center}
\end{table}

The sites are marked, along with some general information, in Figure 1. Dome A is an extended plateau, Dome F is a sharper peak, and Dome C and Ridge B are both nearly level ridges. 

\begin{figure}[htpb]
\plotone{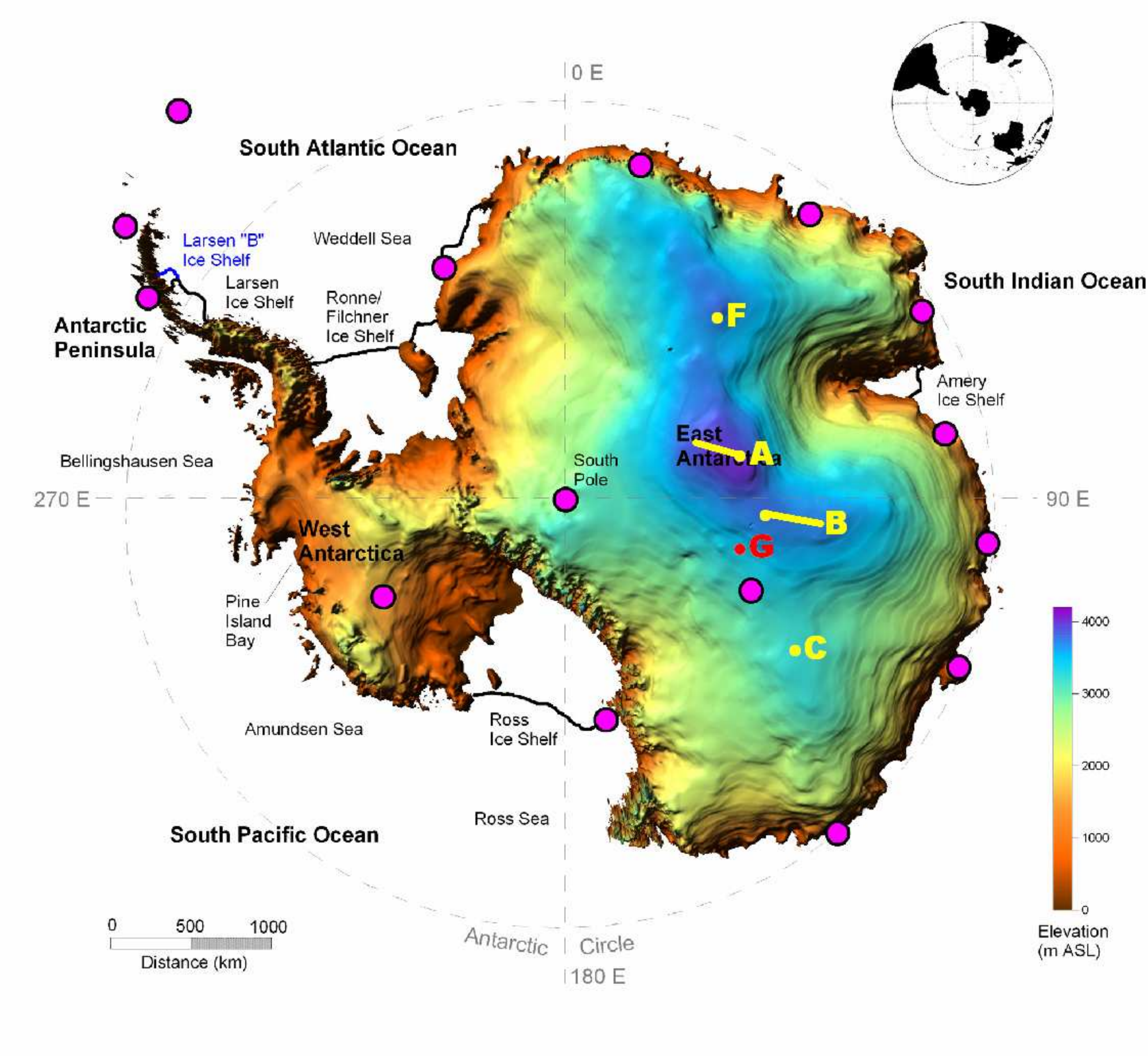}
\caption{Topography of Antarctica, showing the 2010 Geomagnetic Pole (G), the various potential sites (A, B, C, F), and other Antarctic bases.  Adapted from Monaghan and Bromwich (2008), and based on data from Liu \etal (2001).\label{f1}}
\end{figure}

\section{Boundary layer characteristics}
Figure 2 shows the predicted wintertime median boundary layer thickness, from Swain and Gallee (2006a, SG06a). It was this picture that originally suggested to us that the existing bases were not necessarily the best sites. Dome F has marginally the thinnest predicted height at 18.5m; the minimum near Dome A is 21.7m, Ridge B is $<$24m all along its length, while Dome C is 27.7m. Although these differences are small, they have significant implications for the design and cost of any optical/NIR telescope, which must either be above the boundary layer, or fitted with a Ground Layer Adaptive Optics (GLAO) system. Note that these are predicted median values only; Swain and Gallee (2006b, SG06b) predict dramatic and continuous variation of the thickness of the boundary layer at all candidate sites, and this is borne out by actual data from Dome C  (e.g. Aristidi \etal 2009) and Dome A (Bonner \etal, in preparation)

\begin{figure}[htpb]
\epsscale{0.9}
\plotone{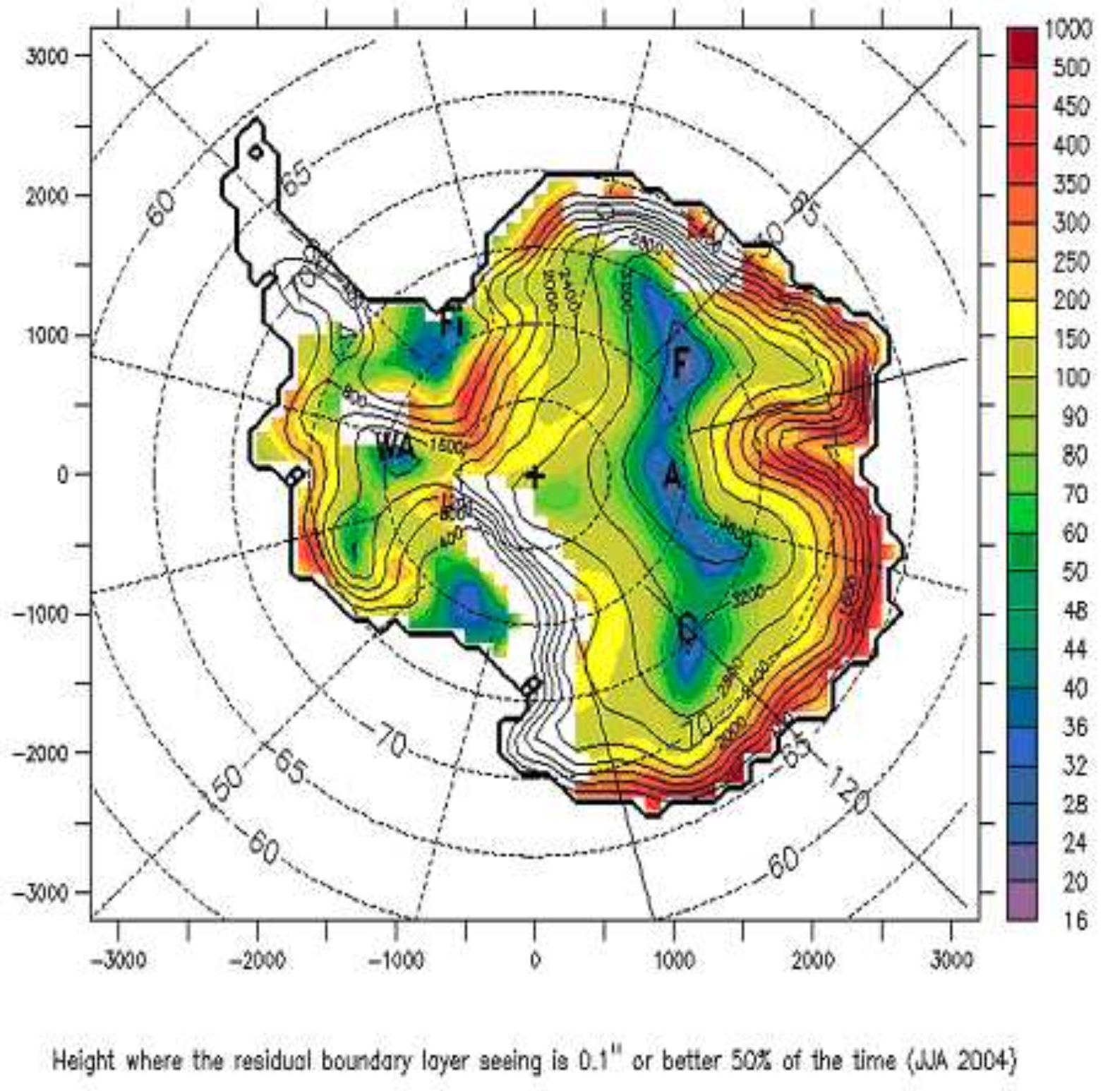}
\caption{Predicted winter (June/July/August) median boundary layer thickness from SG6a. Note the slightly different orientation (105\dg E is horizontal) for all Swain and Gallee plots compared with all others. \label{f2}}
\end{figure}

SG06a also predict that surface seeing is not perfectly correlated with boundary layer thickness, and that the best surface seeing is to be found at Domes C and F. However, it exceeds 1\as even at those sites, so there are no sites in Antarctica with surface seeing as good as the best temperate sites. For any GLAO system, both the thickness of the boundary layer and the surface seeing must be taken into consideration.

\begin{figure}[htpb]
\epsscale{0.9}
\plotone{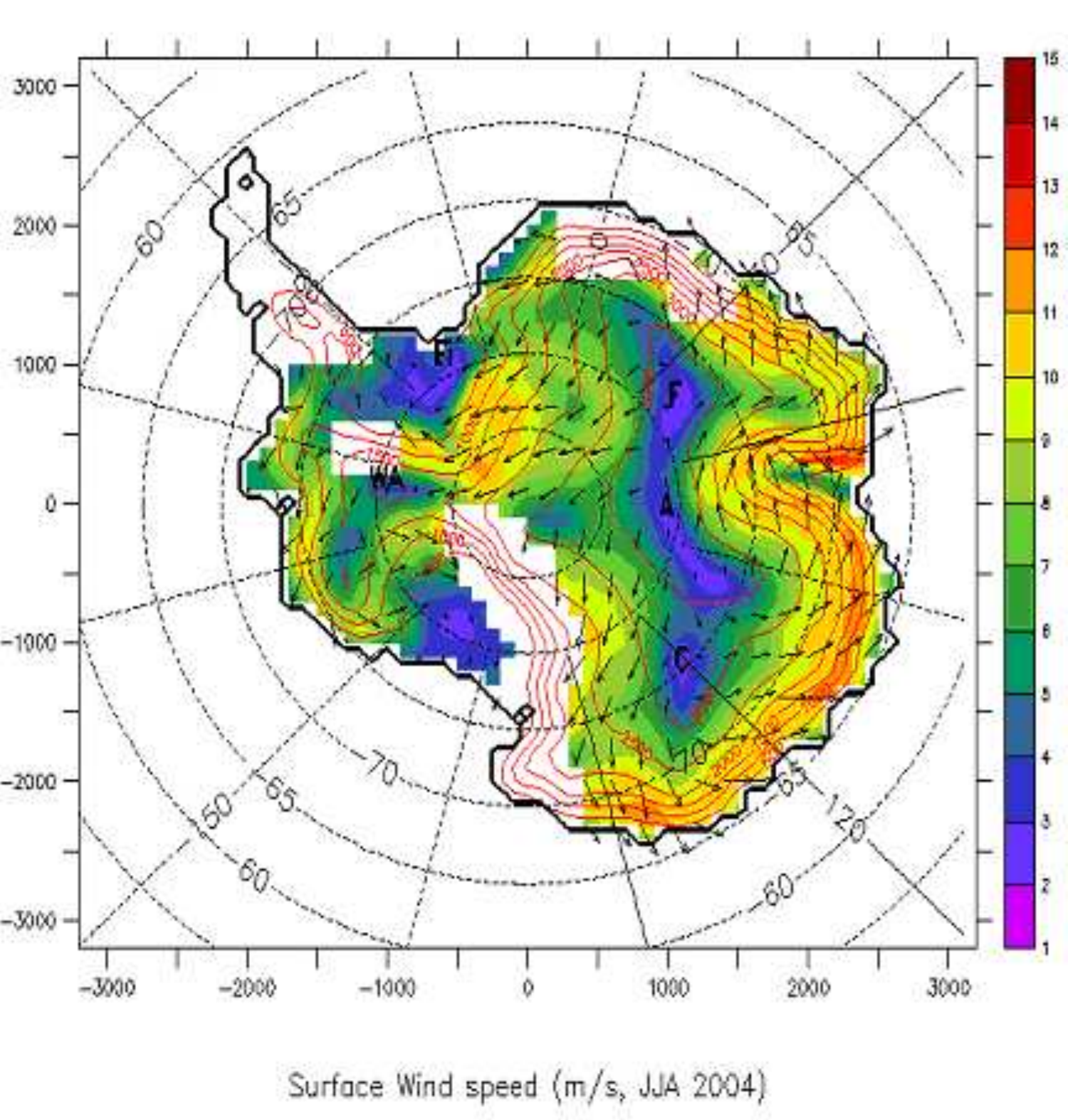}
\caption{Average winter surface wind velocity and speed, from Swain and Gallee (2006b). \label{fig3}}
\end{figure}

SG06b also estimated the average surface wind speeds (Figure 3), showing essentially identical behaviour. Dome F offers the most quiescent conditions, followed by Dome A/Ridge B, and then Dome C.

Other surface wind speed predictions have been made by van Lipzig \etal (2004) (Figure 4), and Parish and Bromwich (2007)  (Figure 5).

\begin{figure}[htpb]
\epsscale{0.75}
\plotone{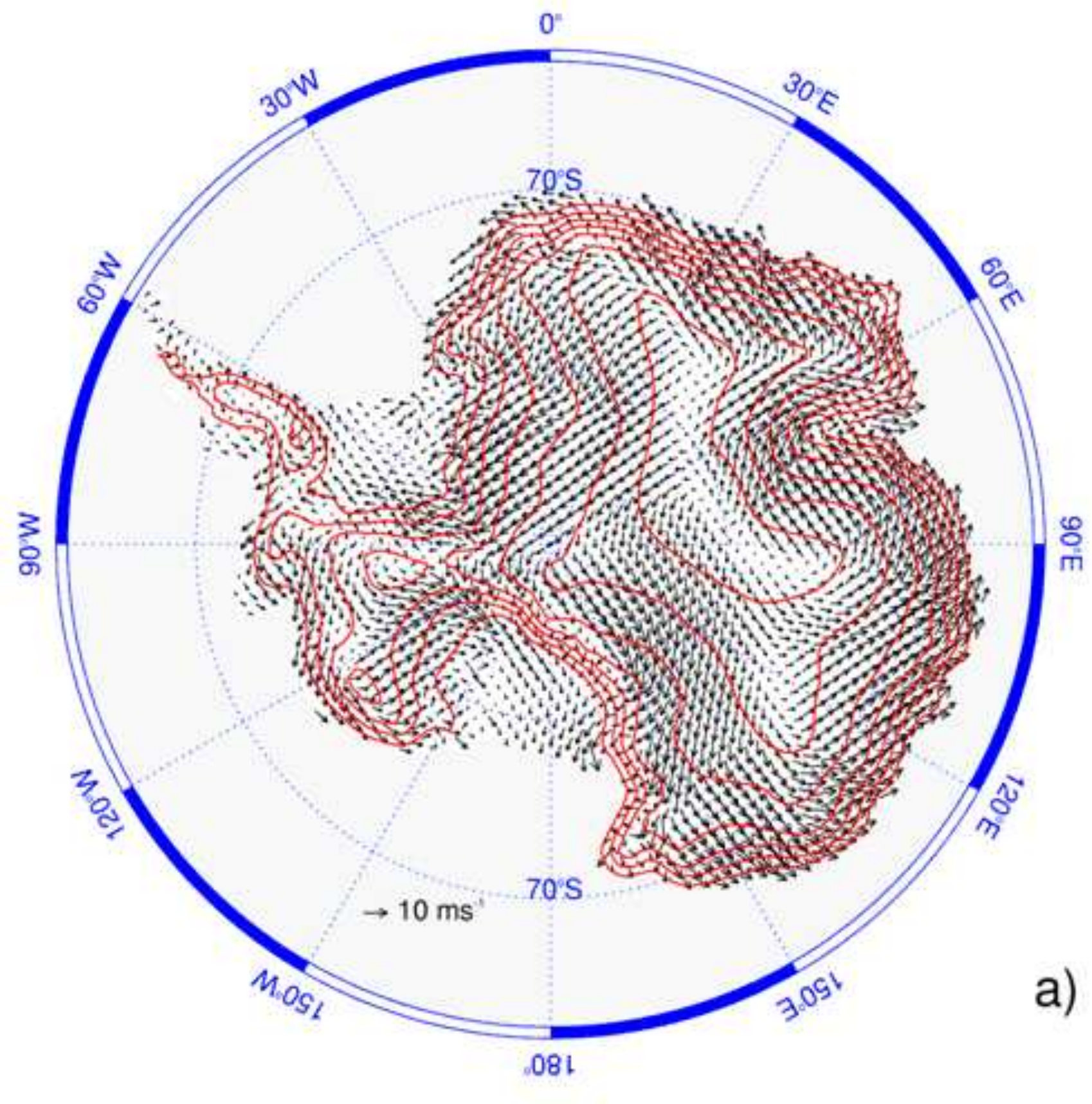}
\caption{Winter surface wind velocities from VanLipzig \etal (2004).  The red contours mark elevation. \label{fig4}}
\end{figure}

\begin{figure}[htpb]
\epsscale{0.75}
\plotone{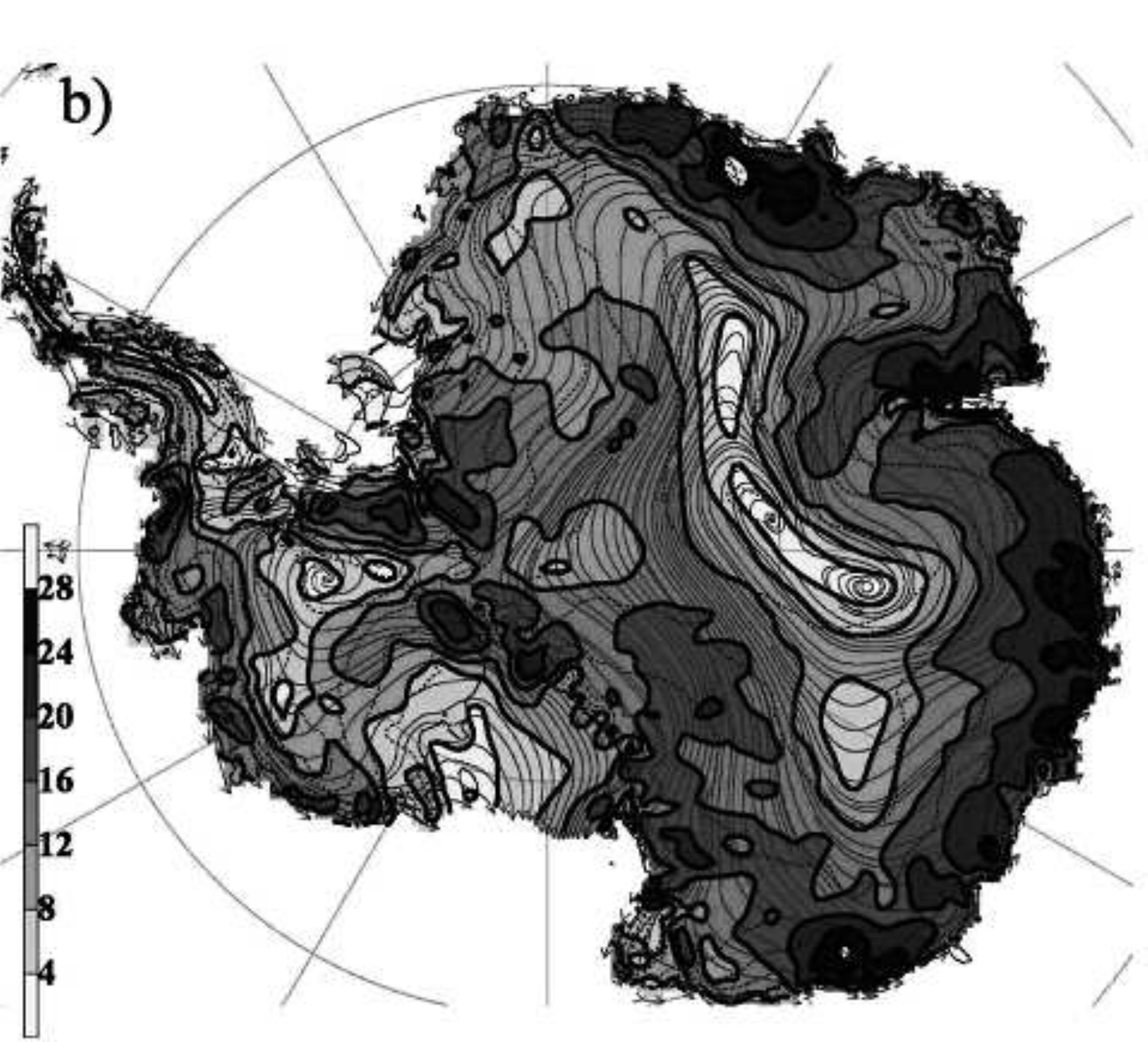}
\caption{Winter wind speed contours at  $\sim$ 100m elevation from Parish and Bromwich (2007). The thin lines are streamlines at $\sim$500m. \label{fig5}}
\end{figure}

There is very good agreement between these three studies. Figure 6 shows an overlay of Figures 2,4 and 5, for the region of interest. The maps are not identical, but the differences are small. In all three maps, there is a clear minimum running from near Dome A through to Ridge B, with an equally good isolated minimum at Dome F. This ‘katabatic ridge’ does not go exactly through Dome A, but is offset towards the South Pole, with minimum at $\sim$81.25\dg S 77\dg E. The deterioration away from this ridge line is very fast: according to SG06a, the predicted boundary layer thickness at Dome A itself is over 30m, i.e. 50\% worse than the minimum, and worse than Dome C. Similarly, along Ridge B, the katabatic ridge is offset from the topographic ridge, in the direction of the lower gradient.

\begin{figure}[htpb]
\epsscale{0.5}
\plotone{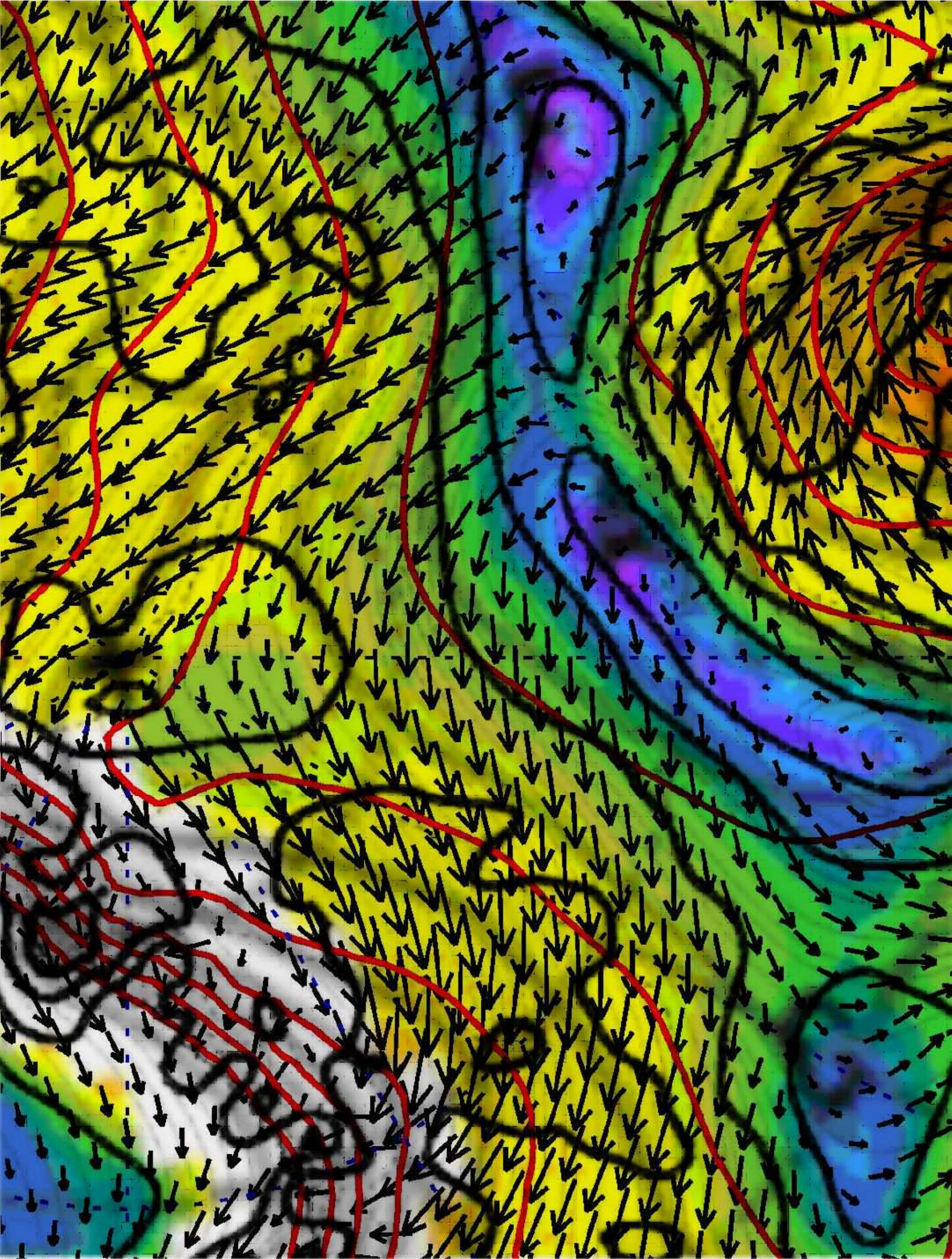}
\caption{Overlay of Figures 2,4 and 5, for the high Antarctic plateau. Red lines are contours. The 80S line of latitude is (just) visible. Near Dome A, there is a minimum in both the predicted wind speed and the predicted boundary layer thickness. However, this minimum is south of Dome A itself at (80.37\dg S, 77.53\dg E) in all three studies.  \label{fig6}}
\end{figure}

\section{Cloud cover} 
The only long-term,comparative observations of cloud cover for the sites under consideration are from passive satellite measurements. Figure 7 shows the average seasonal cloud cover maps for Antarctica, for the period July 2002-July 2007, from an analysis of Aqua MODerate-resolution Imaging Spectroradiometer (MODIS) imagery by the Clouds and Earth’s Radiant Energy Experiment (CERES) using the methods of Minnis \etal (2008) and Trepte \etal (2002). The least cloud cover occurs during the winter, averaging about 0.2 for all the sites, while the most occurs during the summer, averaging about 0.5. in any given season, the cloudiness of the marked sites (Domes A,B,C,F) is very similar, but the least cloud cover is generally between Dome C and Ridge B. 

Data from the Ice, Cloud, and Elevation Satellite Geoscience Laser Altimeter System (GLAS) during October 2003 show a similar range in that area (Spinhirne \etal 2005). The nighttime cloud cover from the GLAS GLA09 V028 5-Hz, 532-nm product for 18th September -- 11th November 2003, plotted in Fig. 8a, shows less structure, but a similar range of values over the highest areas. The resolution of the GLAS data has been decreased to reduce the noisiness of the plots. The relative-maximum “ring” of CERES-MODIS cloudiness (Fig. 8b), seaward of the highest altitudes in eastern Antarctica, is absent in the GLAS data and the daytime CERES-MODIS cloudiness (not shown). This relative maximum artifact is apparently the result of colder-than-expected air at lower elevations during the night. For the extremely cold Antarctic surfaces, the CERES-MODIS cloud detection relies almost entirely on a single infrared temperature threshold at night and will miss clear areas when the actual and temperature is significantly less than its predicted counterpart. Nevertheless, the nighttime Aqua CERES-MODIS average for each of the sites is within 0.04 of the corresponding GLAS 1\dg values for the same period.

\begin{figure}[htpb]
\epsscale{1.0}
\plotone{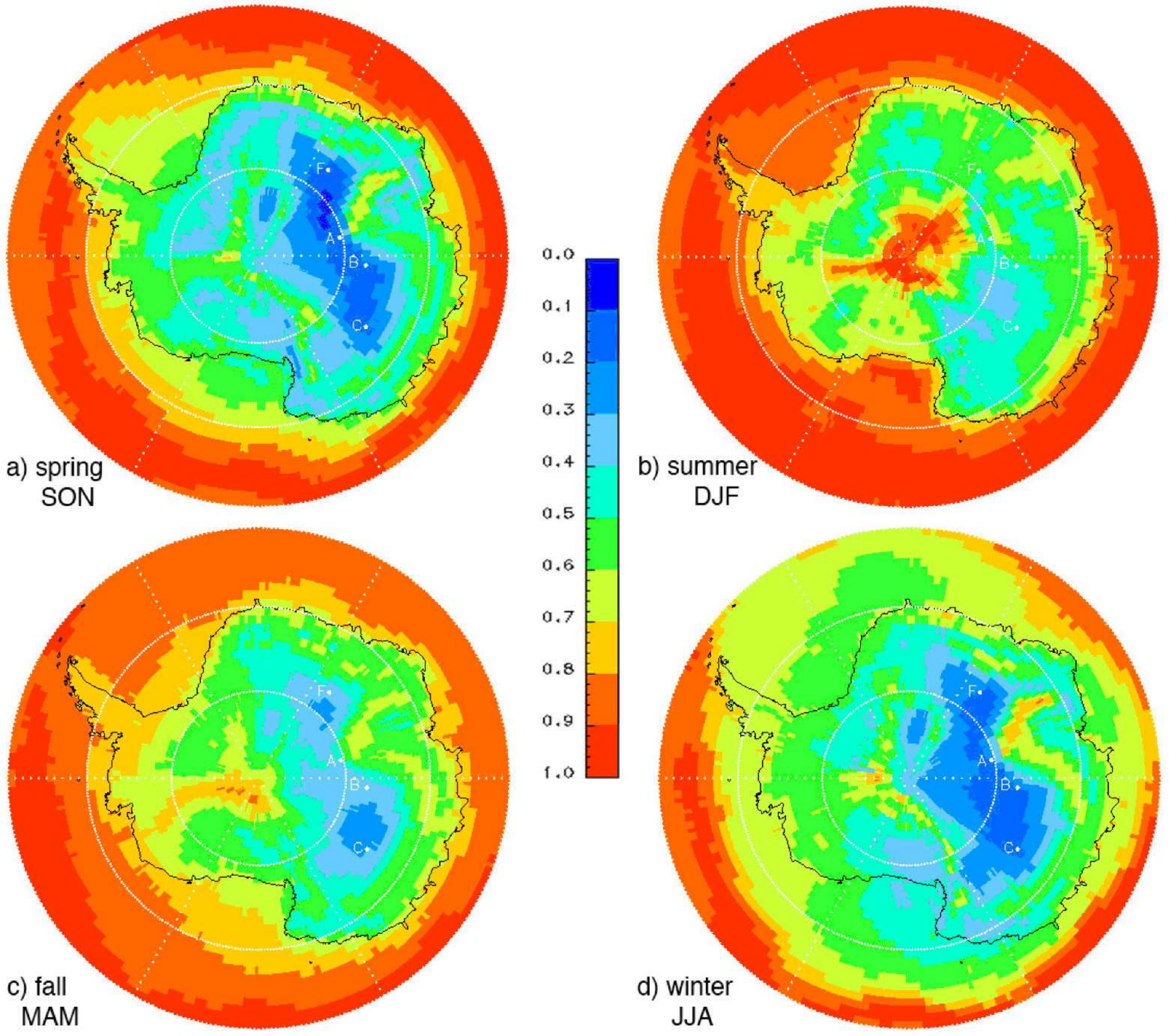}
\caption{Average seasonal cloud cover for the years July 2002- July 2007 from Aqua CERES-MODIS results. SON refers to Sept/Oct/Nov, etc. \label{fig7}}
\end{figure}

\begin{figure}[htpb]
\epsscale{1.0}
\plotone{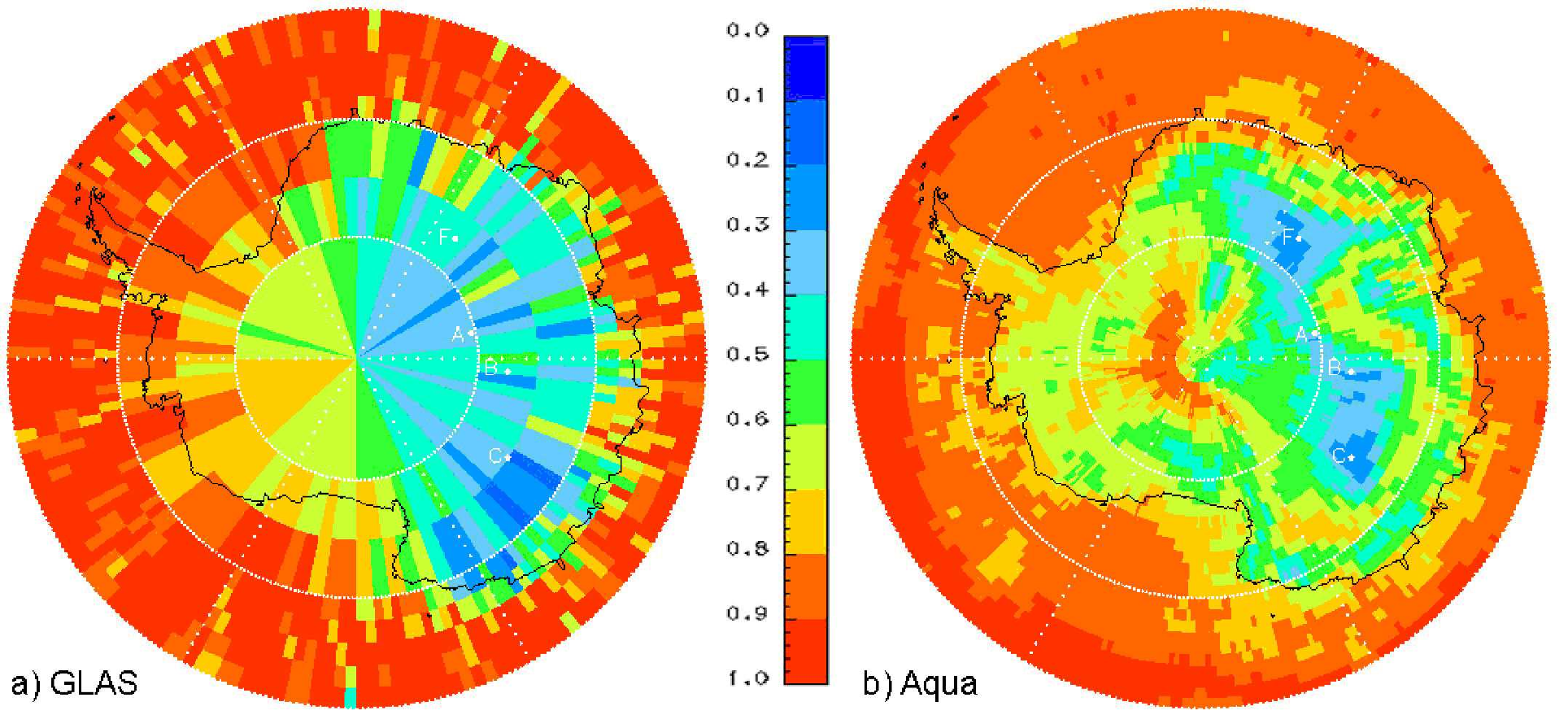}
\caption{Nocturnal fractional cloud cover from satellite instruments for 18 September – 11 November 2003. (a) GLAS, (b) Aqua CERES-MODIS. \label{fig8}}
\end{figure}

\begin{figure}[htpb]
\epsscale{1.0}
\plotone{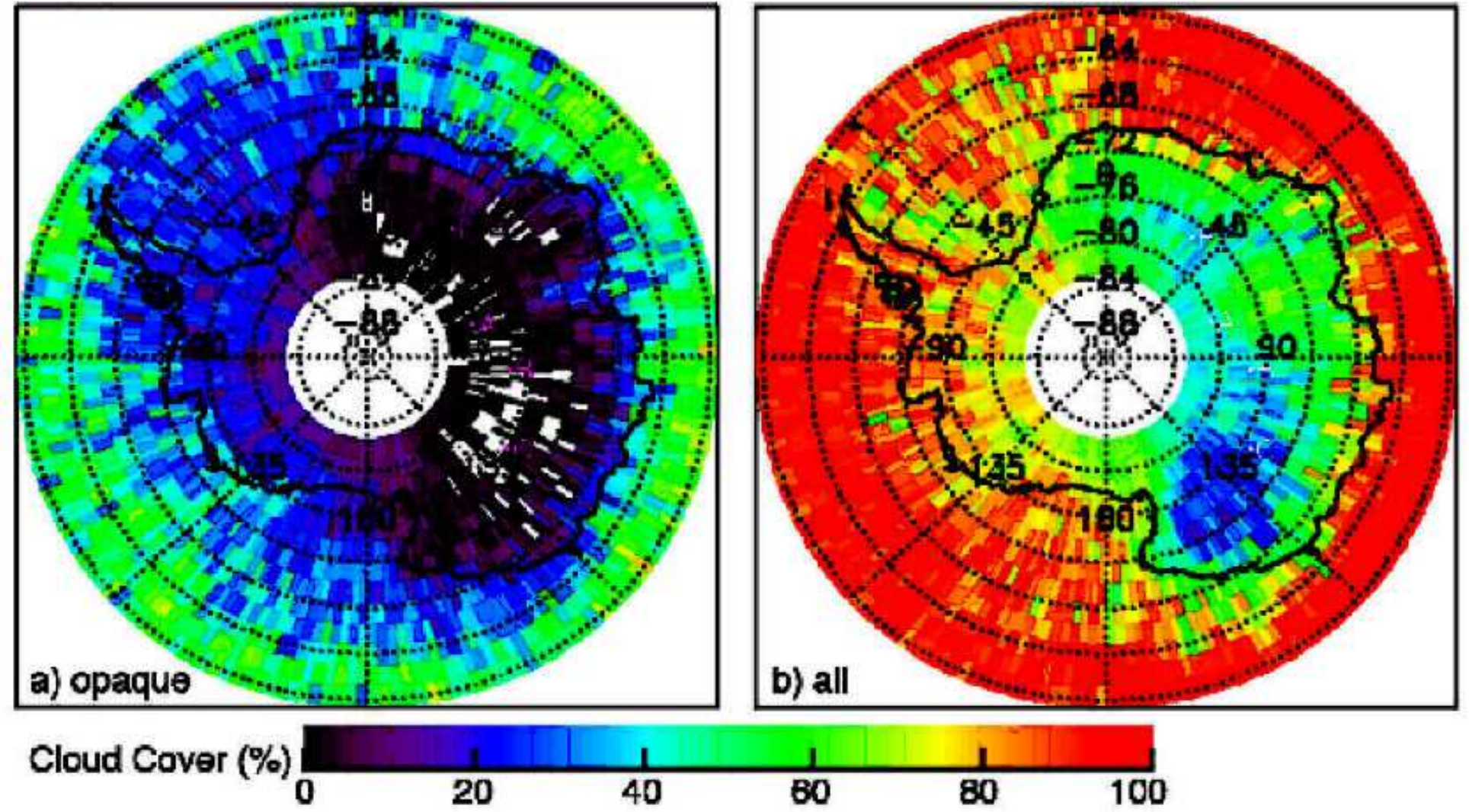}
\caption{Mean cloud amounts derived from CALIPSO lidar data, June – August 2007. The scattered white areas in the left-hand plot indicate no detected opaque cloud cover. \label{fig9}}
\end{figure}

\begin{table}[htpb]
\begin{center}
\caption{Average percentage of nighttime cloud cover for 2002-2007 from Aqua CERES-MODIS cloud products. \label{tbl-2}}
\begin{tabular}{lrrr}
\tableline\tableline
Site& May-Oct&Annual&Summer Range\\
\tableline
Dome A&23.1&24.8&12.5 \\
Dome C&21.5&22.1&9.3 \\
Dome F&25.1&26.5&11.1 \\
Ridge B&24.5&25.7&7.6 \\

\tableline
\end{tabular}
\end{center}
\end{table}

Lidar profiles can provide more accurate estimates of the cloudiness, but satellite-based lidar measurements are limited to only a few years. Mean opaque and total cloud amounts determined from lidar measurements taken from the Cloud-Aerosol Lidar and Infrared Pathfinder Satellite Observations (CALIPSO) satellite (Winker \etal 2007) are shown in Fig. 9 for austral winter 2007. The opaque cloud cover (Fig.9a) corresponds to clouds with optical depths greater than 3.  The opaque cloudiness was nearly non-existent over all of the sites while the smallest total cloudiness (Fig. 9b) occurred over Dome C. At the other 3 sites, the CALIPSO cloudiness tends to be 5-10\% greater than the long-term mean seen in the Aqua results (Fig. 7d). Both datasets have a relative maximum just north of Dome A. The greater CALIPSO cloud amounts are likely due to the limitations of detecting exceptionally thin clouds with passive infrared imagery and interannual variations in regional cloudiness. The Aqua 
cloud data may somewhat underestimate total cloudiness because few clouds having optical thicknesses less than about 0.3 can be detected. Overall, the comparisons with CALIPSO and GLAS indicate that the Aqua cloud cover is within 5-10\% of that detectable with lidars. The cloudiness that is missed by Aqua is very thin, optically speaking, and, therefore, may not have a significant impact.

Annual and May-October average cloud amounts for the 1\dg latitude $\times$ 1\dg longitude regions that include Domes A, C and F are given in Table 2 for 2002-2006. The table also includes the average cloudiness for a 4\dg $\times$ 4\dg region encompassing Ridge B and the summer mean range for all locations. This latter parameter is the difference between the maximum and minimum mean cloud cover for June, July, and August. The cloudiness over the area around Dome C is similar to that observed from the ground for single seasons (20-25\%, e.g. Ashley \etal 2003, Mosser and Aristidi 2007, Moore \etal 2008). Given the comparisons with the surface, GLAS, and CALIPSO observations, it is concluded that the Aqua CERES-MODIS results provide a good representation of the cloud cover over the sites of interest. From Table 2, it is clear that there is little difference in cloud cover among the sites. While Dome C has the least cloud cover, it is only ~4\% less than the maximum at Dome F. Of the three sites having the 1\dg resolution data, Dome C has the smallest interannual range in summer cloudiness. The smallest range, over Ridge B, may be due to its larger spatial domain. The greatest range in mean summer cloud cover is over Dome A, which is not very far from a relative maximum in the total cloud cover. 

Figure 10 shows a closeup of Figure 7(d), the Aqua/\\MODIS winter cloud cover over the high plateau. The ridge of best weather does not to go through Dome A, but is offset by 1-2\dg to the south. This behaviour is seen throughout the year in the Aqua/MODIS data, though it is not present in the CALIPSO data.

\begin{figure}[htpb]
\epsscale{0.5}
\plotone{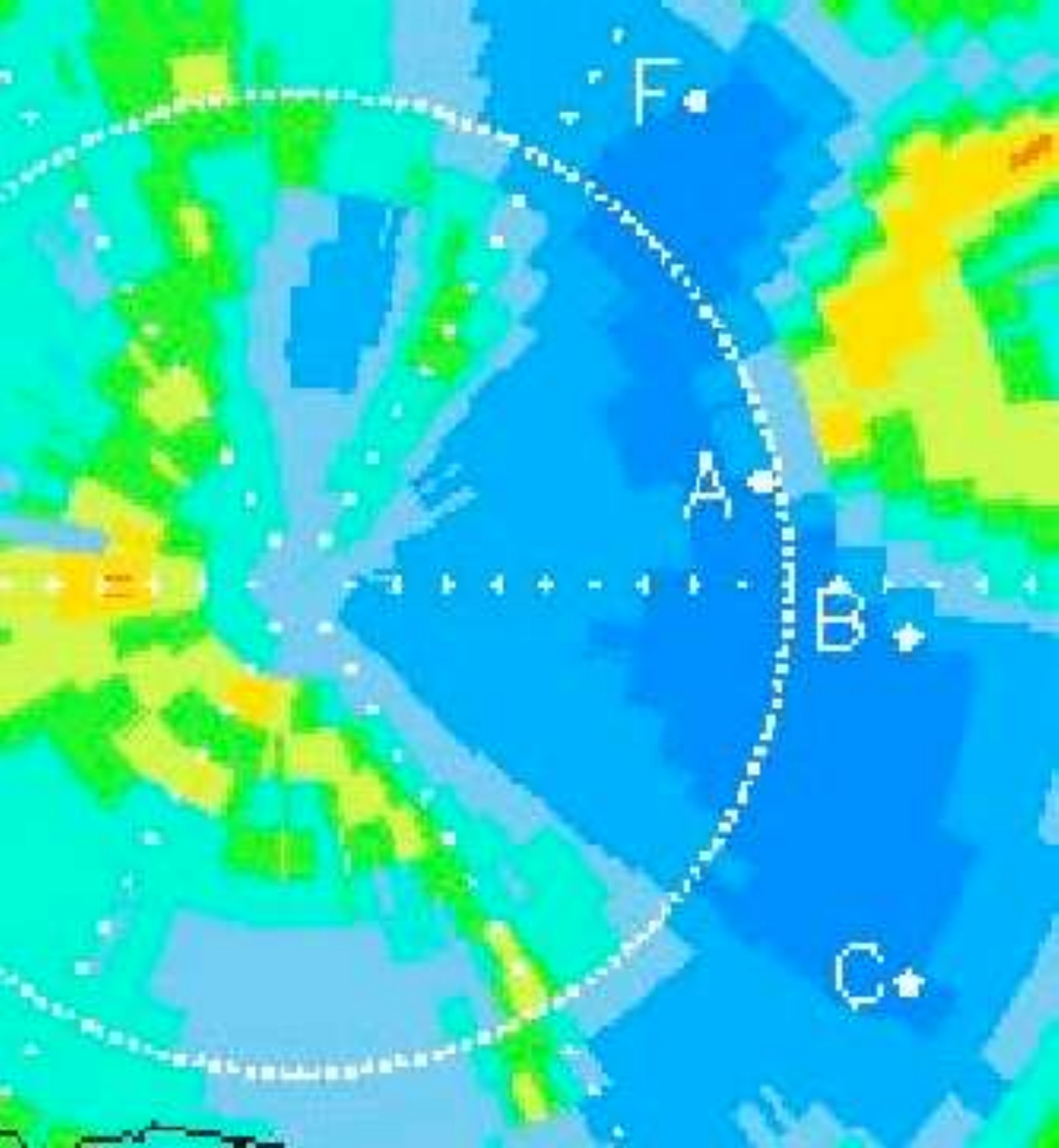}
\caption{Closeup of Figure 7(d), showing the wintertime fractional cloud cover from the Aqua CERES-MODIS results. The ridgeline of minimum cloud cover passes to the south of Dome A. \label{fig10}}
\end{figure}

\section{Aurorae}
Auroral activity depends on solar activity, geomagnetic latitude ($\Lambda$), and magnetic local time (MLT). The relative colours of typical aurorae and the night sky, are such that aurorae brighten the sky most in $U,B,V$ bands, in that order. At temperate latitudes, auroral activity makes a negligible contribution to sky brightness (e.g. Benn and Ellison 2007). The strongest aurorae occur in an oval between 60\dg$<|\Lambda|<$75\dg, stronger in the direction MLT=12. Inside this oval, there is a region of lower activity, still at a level much higher than at temperate latitudes (e.g. Hardy \etal 1985). The activity levels at the magnetic poles themselves have no measured dependence on solar activity level, or of course on MLT (Hardy  \etal 1991, H91). Daily Antarctic auroral activity is shown online \footnote{\tt http://www.swpc.noaxa.gov/pmap/pmapS.html}. 

The Geomagnetic South Pole is currently near (80\dg S 109\dg E), and moving too slowly to matter even on the timescales of Antarctic astronomy \footnote{see \tt http://www.geomag.bgs.ac.uk/poles.html}. The geomagnetic latitude of the sites is given in Table 3. Dome A, Dome C and Ridge B are all within 6\dg of the geomagnetic pole, while the South Pole and Dome F are on the edge of the auroral oval at 10\dg and 13\dg respectively.

The effect of auroral activity on sky brightness for Antarctic sites was investigated in detail by Dempsey, Storey and Phillips (2005, D95). They determined that, at the South Pole, auroral emission was a significant, but not catastrophic, issue for sky-limited optical astronomy in $B,V$ and $R$ bands. They estimated auroral contributions to the sky brightness of $21.7 - 22.5^m$ /arcsec$^2$ in $B$ band, and $21.8 -– 22.5^m$/arcsec$^2$ in $V$ band. These compare with dark sky brightness values at good temperate sites of $22.5 - 23.0^m$/arcsec$^2$ in $B$ band, and $21.5 -– 22.0^m$/arcsec$^2$ in $V$. They found that the aurora were brighter at solar {\it minimum}.

Unfortunately, the D95 paper contains an error in the geomagnetic latitude of all the sites considered, so the extrapolation of the South Pole result to other sites is incorrect. We have repeated the exercise in that paper, of using the auroral models of H91 to predict the auroral contribution at other sites. H91 give average solar electron flux intensity maps as a function of solar activity level $K_p$, ranging from 1 (low) to 6 (high), geomagnetic latitude, and local time. We have integrated these models over local time (using the logarithmic average), to find the approximate median contribution as a function of geomagnetic latitude and solar activity level (Figure 11). It is striking that for $| \Lambda | > 77$\dg, the auroral flux is {\it anti-correlated} with solar activity level, consistent in both sense and magnitude with that seen by D95. As we go closer to the geomagnetic pole, both the overall flux level and its variability are reduced. The maximum and minimum levels are shown, for all the sites under consideration, in Table 3.

\begin{figure}[htpb]
\epsscale{1.0}
\plotone{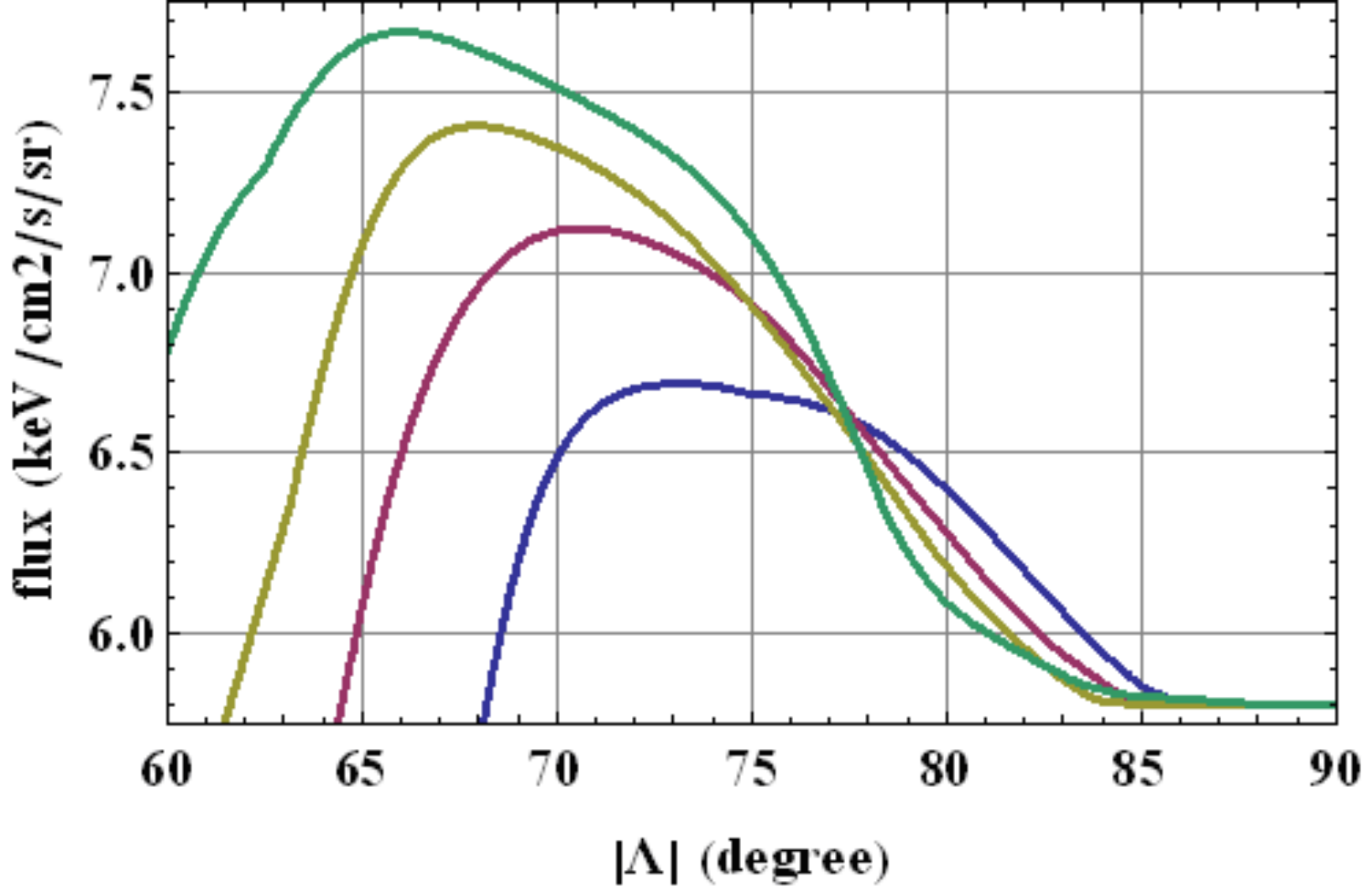}
\caption{Average integral energy flux as a function of solar activity level and geomagnetic latitude $\Lambda$. Activity levels plotted are, in order of lowest peak to highest, $K_p=0$ (blue), 2 (red), 4 (orange), 6 (green). \label{fig11}}
\end{figure}

\begin{table}[htpb] 
\begin{center}
\caption{Relative auroral energy flux and sky brightness contributions ($AB$ mags /arcsec$^2$) at the various sites, at minimum and maximum solar activity levels. Flux is as compared with the average at the geomagnetic pole, which is $6.31 \times 10^5  keV/sr/s/cm^2$ , independent of MLT or solar activity (Hardy \etal 1991). \label{tbl-3}}

\begin{tabular}{|p{0.25in}|p{0.4in}|p{0.2in}p{0.2in}p{0.2in}|p{0.2in}p{0.2in}p{0.2in}|}
\tableline\tableline
Site &	$\Lambda$ & &  $K_p$=0  & & &  $K_p$=6 & \\
& 2010& Flux &   $\mu_B$   &  $\mu_V$	&  Flux    & $\mu_B$ & $\mu_V$ \\
\tableline
SP &	-80\dg&	3.98&  21.7&  21.8&	1.92&  22.5&  22.5\\
DA&	-84.9\dg&	1.16&  23.0&  23.1&	1.06&  23.1&  23.1\\
DC&	-84.1\dg&	1.37&  22.9&  23.0&	1.10&  23.1 & 23.1\\
DF&	-77.0\dg&	6.58&  21.2&  21.3&	8.12&  20.9&  20.9\\
DB	&-85.4\dg& 1.08&  23.1 & 23.2 &	1.06 & 23.1 & 23.1\\
RB & -87.1\dg& 1.00 & 23.2 & 23.3&1.02 & 23.2 & 23.2 \\
\tableline
\end{tabular}
\end{center}
\end{table}

We find that Dome A, Dome C and Ridge B all have remarkably similar and constant average auroral contribution to the sky brightness, at a level $\sim23^m/$arcsec$^2$, at all times and in both $B$ and $V$ bands. This corresponds to an increase in sky brightness compared with the best temperate sites of almost a factor of 2 at $B$ and about 20-30\% at $V$. The difference between these three sites is small, though Dome C is marginally the worst. The optical sky brightness at Dome F is dominated by aurorae, most of the time.

\section{Airglow}
Airglow emission from OI is responsible for very strong emission features at 557.7nm and 630nm, NO$_2$ is responsible for a 500--650nm continuum, while OH dominates the night-sky brightness from 700nm -- 2300nm. Airglow was considered by Kenyon and Storey (2006), who found no strong evidence for large systematic variations in airglow emission as compared with temperate sites. Recently, more detailed predictions have been made by Liu \etal (2008), for OI and OH emission, for all latitudes, seasons and local times. The predicted average emission for 20hr-4hr local time, is shown in Figures 12(a) (OI) and 12(b) (OH). Results for other times are not available, but there is almost no time dependence for Antarctic winter emission for 20-4h, and it seems reasonable to assume that the maps apply for all 24 hours. The models are validated against data from the WINDII satellite, but this is for temperate latitudes only, so the predictions for polar regions are unverified.

 There are several features to note: the predicted OI emission is very strong in Antarctica in winter, almost an order of magnitude greater than at temperate sites. In principle, this can be filtered out with narrow-band filters. For OH emission, the Antarctic winter values are $\sim$ 30\% higher than temperate sites. However, the model predicts a striking `OH hole' over Antarctica each October, with OH emission 6 times less than at temperate sites. The hole is predicted to persist all summer, but sadly decays in Autumn just as soon as there is any dark time to use it. 

 Direct OH emission measurements from the South Pole \footnote{\tt cedarweb.hao.ucar.edu/wiki/index.php/Instruments:spm} provide some partial support for this prediction, with the OH emission routinely settling down to levels $\sim$5 times lower than the winter median, for periods of several days at a time. Unfortunately, comparative data for temperate sites is not currently available.

If confirmed, this `OH hole' would be a striking extra advantage for Antarctic astronomy, since fields could be observed in $J$ and $H$ bands to a depth comparable with \Kdark. It happens that the hole coincides with the best accessibility for the South Galactic Pole and Chandra Deep Field South. The amount of dark time available at those times of year is very limited: assuming a required solar elevation below $-10$\dg, there is $\sim$3.5 hours/night of dark time at Dome C at the equinoxes, and none at all at Dome A. So although the depth of the hole is greatest at the Pole, our ability to make use of it depends on being as far north as possible, and in this respect Dome C has a distinct advantage compared with the other sites.

The OH and OI data are too coarse (5\dg latitude bins) to make very useful predictions for the average emission values at the various sites; but the model predicts higher OI emission, and lower OH emission as we approach the Pole.

\begin{figure}[htpb]
\epsscale{1.0}
\plotone{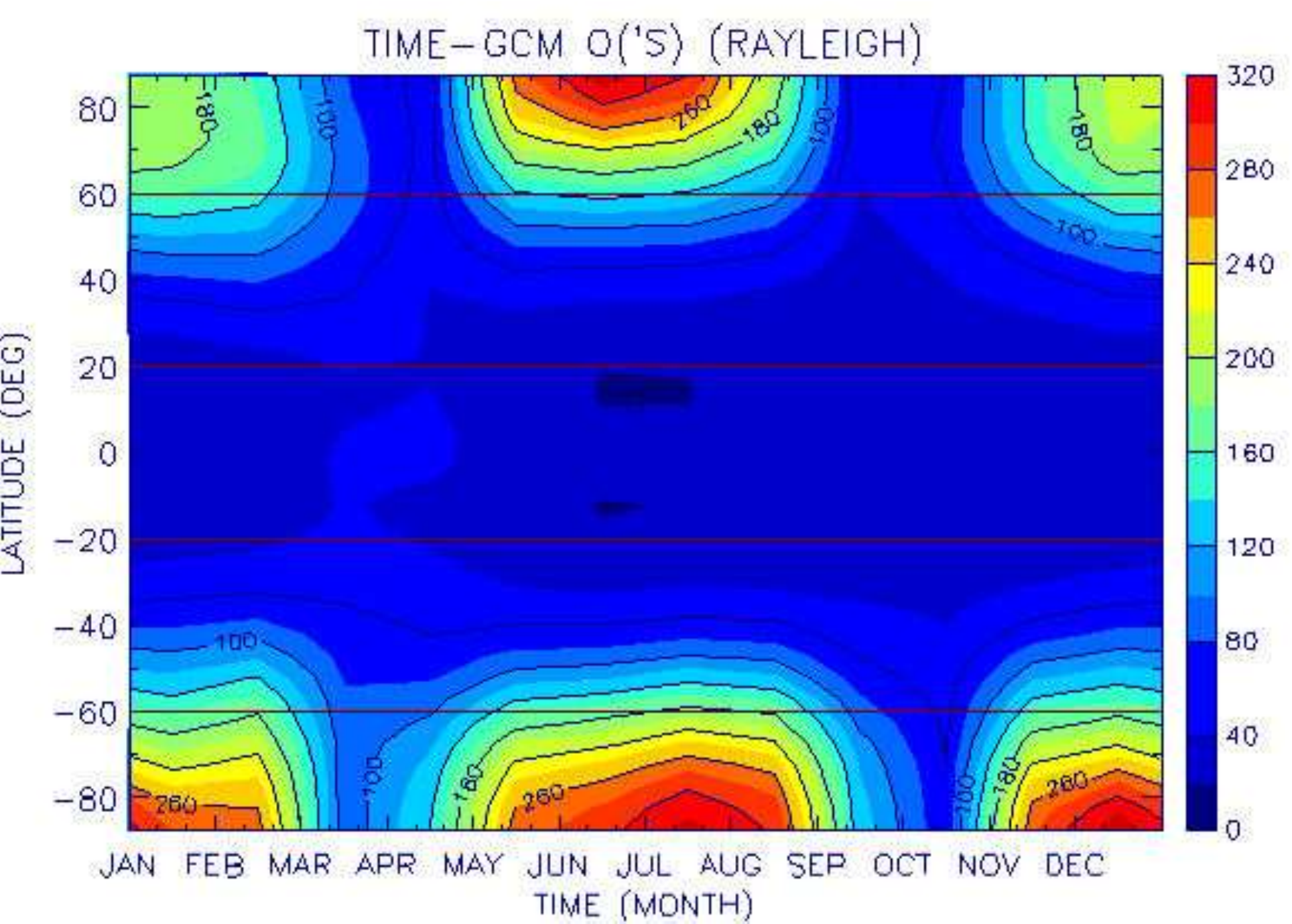}
\plotone{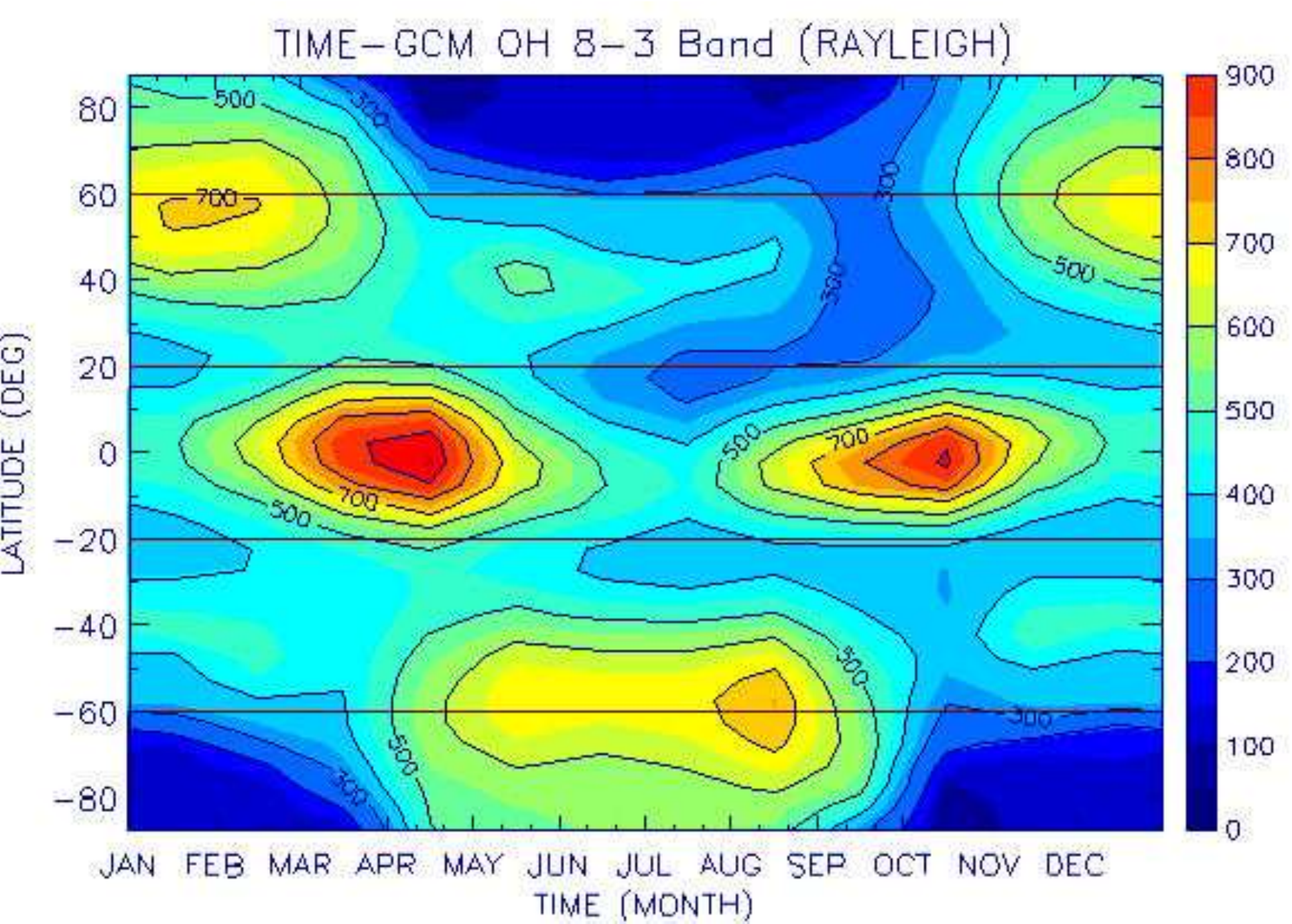}
\caption{Predicted average OI 557.7nm (upper plot) and OH (8-3) (lower plot) emission as a function of latitude and month, from the models of Liu et al (2008). Units are Rayleighs. \label{fig12}}
\end{figure}

\section{Precipitable water vapour}
SG06b produced a map of predicted average PWV (Figure 13). They predict Dome A to be the best existing site; Dome F to be very nearly as good; Dome C about a factor of two worse, with Ridge B intermediate. The best location of all is once again between South Pole, Dome A, and Dome F.

\begin{figure}[htpb]
\epsscale{0.9}
\plotone{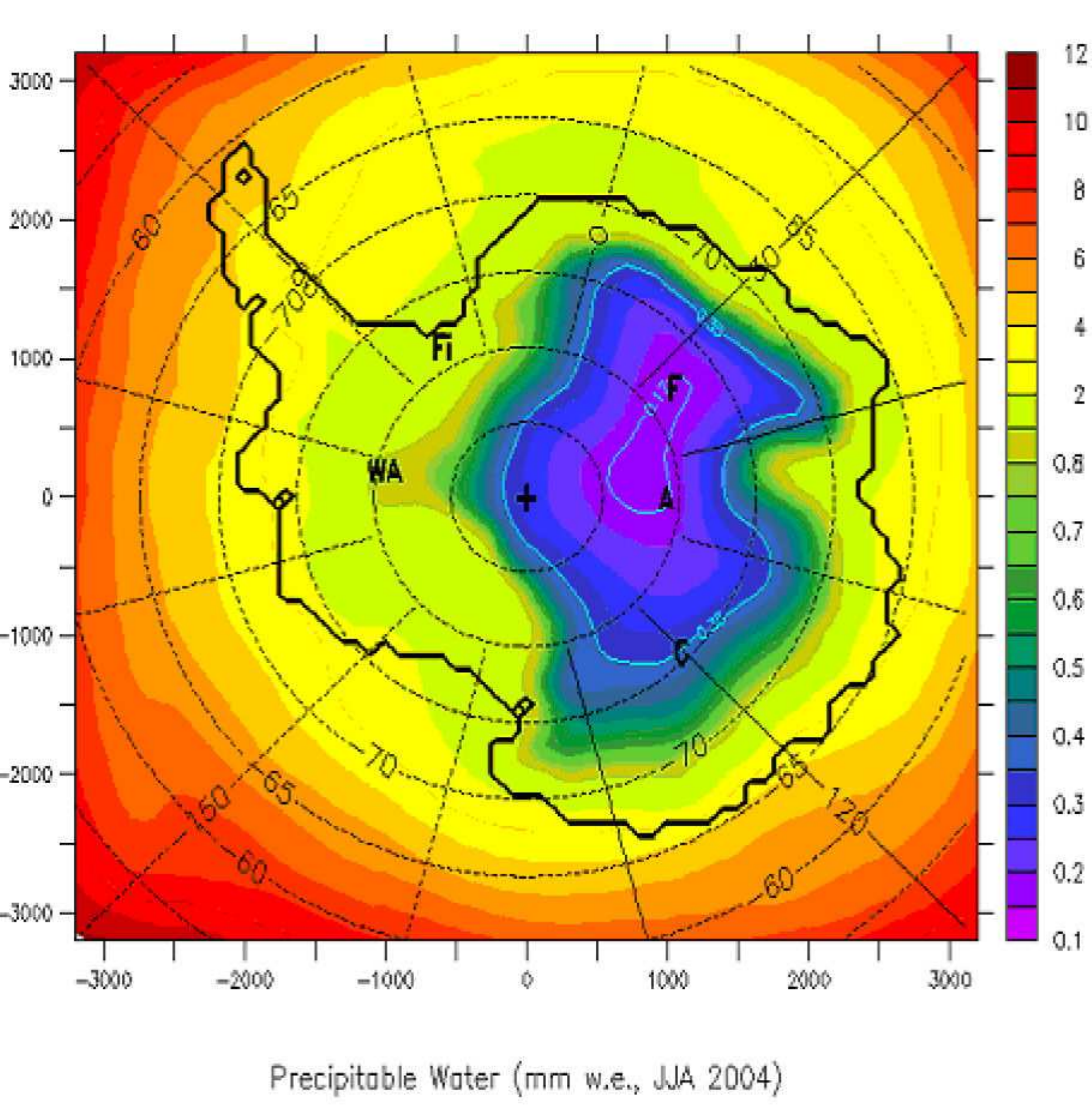}
\caption{Predicted winter PWV from Swain and Gallee (2006b) \label{fig18}}
\end{figure}

The MHS experiment on the NOAA-18 satellite allows estimation of the daily PWV directly for the whole of Antarctica. The estimate has been validated against ground based measurements at Dome A (Kulesa \etal, in preparation). The statistics for the various sites are shown in Table 4. The agreement with the SG06b predictions is quite good: the difference between Dome A and Dome C is not quite as large as predicted, Dome F is a little worse than predicted, while Dome B is a little better, making it as good as Dome F. Ridge A again emerges as a significantly better site than Dome A.

\begin{table}[htpb]
\begin{center}
\caption{PWV quantiles from the MHS sensor, for 2008. Units are microns. `Winter' means days 120-300.  \label{tbl-4}}
\begin{tabular}{lrrrrrrr}
\tableline\tableline
              &  SP &       DC  &      DA &       RA   &     DB  &      DF \\
\tableline
Annual med.  &     437  &     342  &      233  &      210  &      274  &      279 \\
Winter med.  &      324 &       235  &      141  &      118   &     163   &     163 \\
Winter 25\% &       258 &       146 &       103 &       77  &      115 &      114 \\
Winter 10\% &       203 &       113  &      71  &      45   &     83    &    90 \\
Winter $\sigma$ &       133 &       122 &       65 &       64 &       67  &      98 \\
\tableline
\end{tabular}
\end{center}
\end{table}

\section{Atmospheric thermal emission}
The atmospheric thermal emission is determined both by the total mass of each atmospheric component above the site, and its temperature profile. For CO$_2$, the mass is proportional to the surface air pressure, which varies from $\sim$575mb at Dome A to $\sim$645mb at Dome C. For water vapour, the temperature profile is paramount, as it limits the saturated mixing fraction. The temperature profile is shown in Figure 14, which shows the winter (May-Aug) mean temperatures from the NCAR/NCEP reanalysis data (Kalnay \etal 1996), as a function of pressure height. The atmospheric emission is dominated by the lower layers. The coldest air is once again between Dome A and South Pole.  Dome F is somewhat better than Ridge B, which is better than Dome C. The difference between Dome A and Dome C is consistently about 3\dg C at all pressure heights.

We have modelled the infrared sky brightness using the Line-by-Line Radiative Transfer Code (see, e.g. Lawrence 2004). For each site the input parameters are the profiles of temperature and pressure, while the relative humidity is set by matching the overall PWV to the winter median from Table 4. The resulting spectra of IR sky brightness are shown in Figure 15. In order of decreasing infrared sky emission, the sites can be ranked as follows: South Pole, Dome C, Dome F, Ridge B, Dome A, and Ridge A. The difference between South Pole and Ridge A is about a factor of 3 in the best thermal infrared windows, but about 1.5 in the optically thick bands that dominate the broad-band sky brightness.

The sky brightness within the \Kdark passband at South Pole is 80-200$\mu Jy$/arcsec$^2$ (Philips \etal 1999). However, there are no measurements elsewhere on the Plateau, and the dominant emission mechanism is unknown, so this value cannot be extrapolated to the other sites. All we can say is that since all the other sites are higher and drier, the background is likely to be lower.

Additionally to any variation of the sky brightness within the \Kdark window, we can expect that the higher and colder sites allow a wider passband; this is because the atmospheric thermal emission defines the redward passband cutoff. From Figure 15, the wavelength where the thermal emission exceeds $\sim100 \mu Jy$ is shifted redward by $\sim0.09\mu m$ between South Pole and Ridge A, an increase in bandwidth of 50\%, and allowing observation out to $\sim 2.5\mu m$.

\begin{figure}[htpb]
\epsscale{1.0}
\plotone{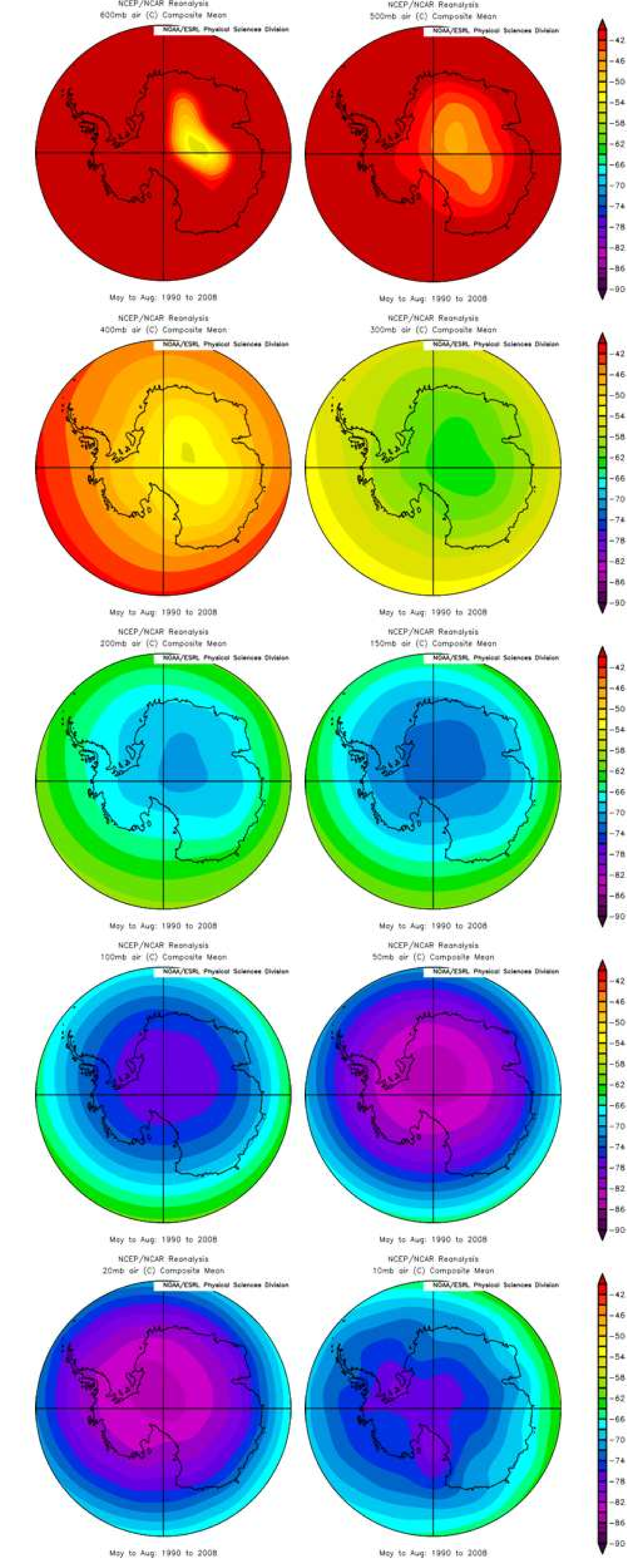}
\caption{Mean winter temperature for 1990-2008, at pressure heights 600, 500, 400, 300, 200, 150, 100, 50, 20, 10mb. Range is -90\dg C (purple) to -40\dg C (red). \label{}}
\end{figure}

\begin{figure}[htpb]
\epsscale{1.1}
\plotone{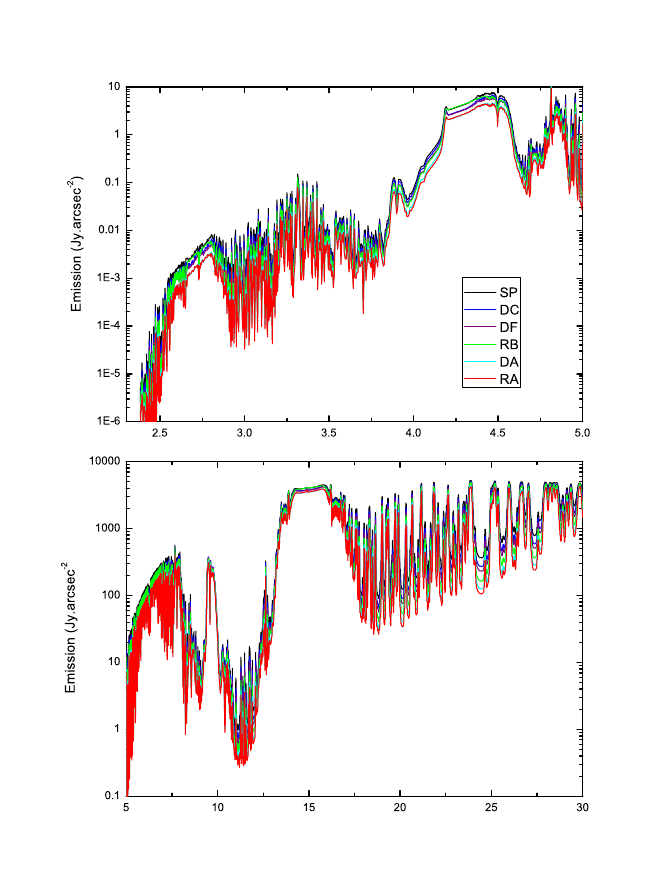}
\caption{ Model infrared sky thermal brightness in $K$,$L$,$M$ bands (top) and $N$ and $Q$ bands (bottom) for various Antarctic plateau sites. Note that OH airglow emission is not included.
 \label{}}
\end{figure}

\section{Surface temperature}
The coldest possible surface temperature implies the lowest telescope emission in the thermal infrared. Figure 16 shows the predicted winter surface temperature from SG06b, while Figure 17 shows the measured average winter surface temperature, derived from the Aqua/MODIS data for 2004-2007. As expected, the ridge along Dome F -- Dome A -- Ridge B defines the coldest regions, with a separate (and almost as cold) minimum at Dome C. The ridge of minimum temperature again misses Dome A, passing through 81-82\dg S. Actual values for the sites are given in Table 5. Ridge B is colder than Dome A, and nearly as cold as anywhere on the plateau; Dome C is also remarkably cold. However, all the sites are within a few degrees of each other, and the effect on overall telescope emission is modest.

\begin{table}[hptb]
\begin{center}
\caption{Average winter (June/July/August) surface temperatures \label{tbl-5}}
\begin{tabular}{lrrr}
\tableline\tableline
Site & Lat & Long & T (K) \\
\tableline
Dome A &80.37S &77.53E & 204.1 \\  
Dome C &75.06S &123.23E & 204.9 \\  
Dome F &77.19S& 39.42E & 204.9 \\  
Dome B &79S &93E & 203.6 \\  
Ridge B &76S &94.5E & 206.9 \\ 
Ridge A &81.5S &73.5E & 203.5 \\  
\tableline
\end{tabular}
\end{center}
\end{table}

\begin{figure}[hptb]
\epsscale{0.9}
\plotone{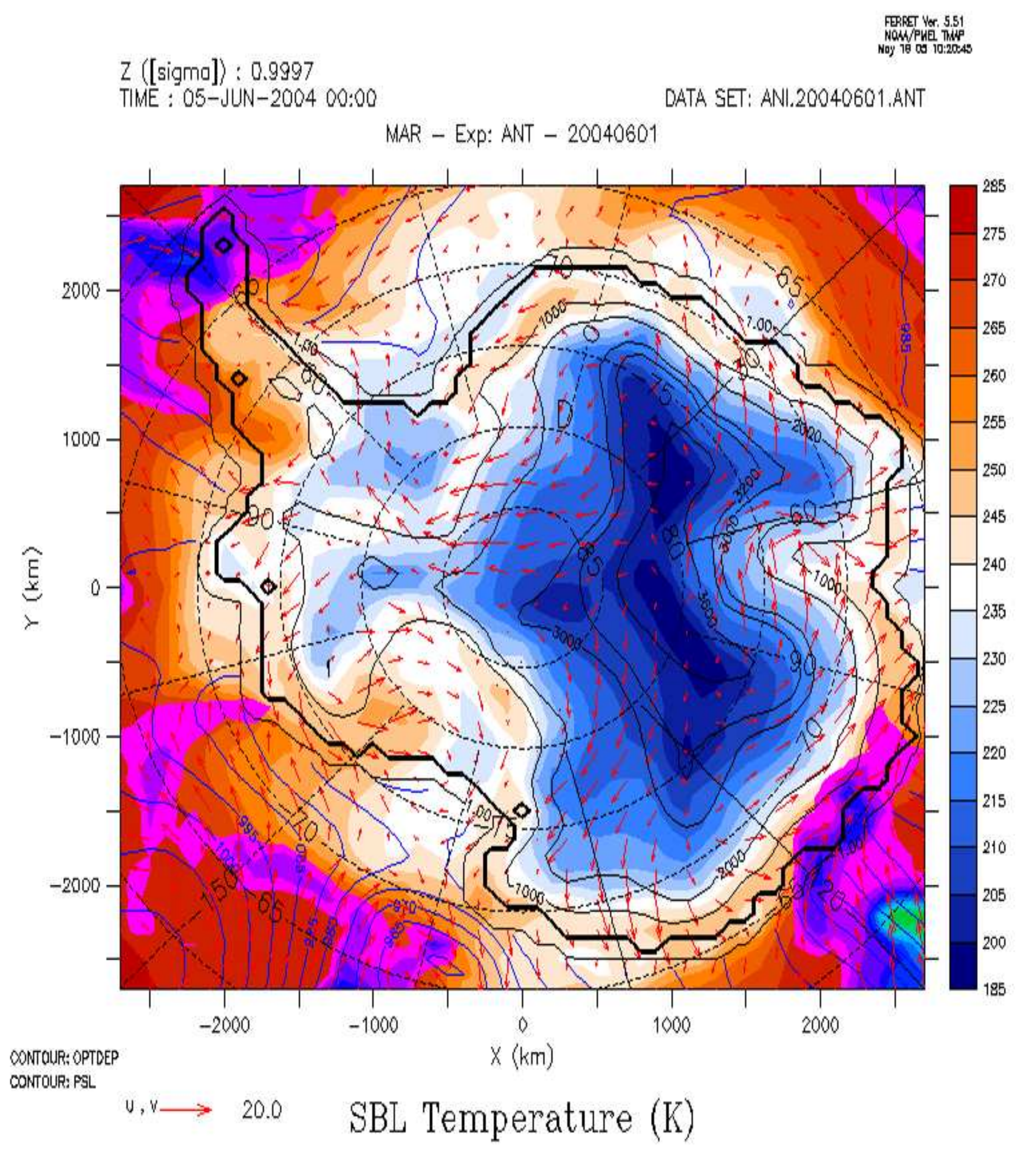}
\caption{ Predicted winter surface air temperature from Swain and Gallee (2006b). \label{fig19}}
\end{figure}

\begin{figure}[hptb]
\epsscale{0.9}
\plotone{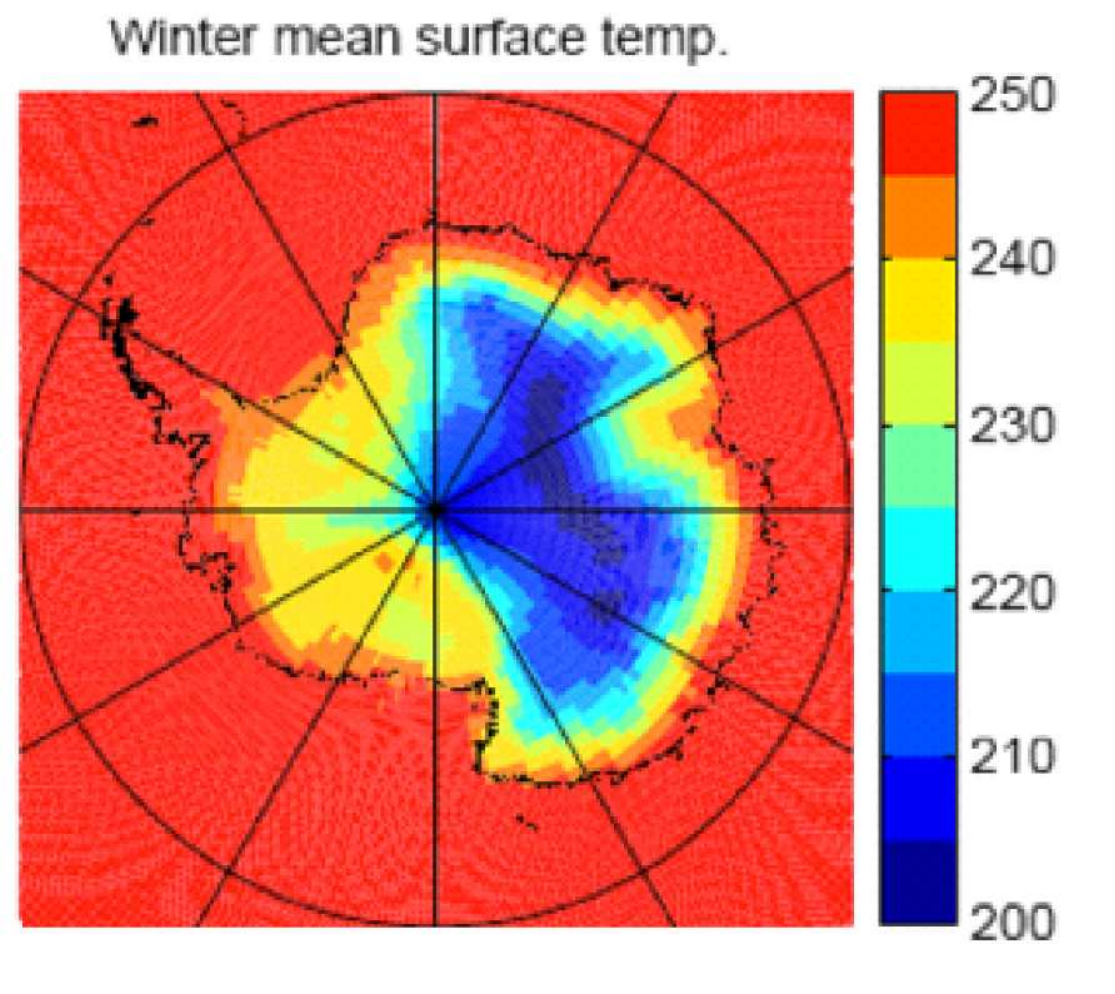}
\caption{Observed winter (June/July/August) temperature from the Aqua/MODIS data for 2004-2007. \label{fig20}}
\end{figure}

\section{Free atmosphere seeing}
We have direct measurements of the free seeing (i.e. above the surface boundary layer) from Dome C, where it is $0.27-0.36''$ (Lawrence \etal 2004, Agabi \etal 2006, Aristidi \etal 2009), and also from South Pole (Marks \etal 1999), where it is $0.37\pm0.07''$.

Estimating the seeing directly from meteorological data is extremely uncertain, because the seeing is in general caused by turbulent layers much thinner than the available height resolution. However, the importance of the free seeing makes it worthwhile to attempt some estimate of its variation between different sites, however crude. 

Figure 18 is taken from the NCAR/NCEP reanalysis data (Kalnay \etal 1996), and shows the mean wintertime (May-Aug inclusive) wind speed, over the years 1979-2008, as a function of pressure height. At all heights, there is a general minimum over the Antarctic plateau. The minimum is rather weak at the lowest elevations, but includes the Dome A -- South Pole region; as we move to higher elevations, the minimum becomes strikingly defined, and very symmetrical around a point half way between Dome A and South Pole. We can be reasonably confident that the best seeing will be associated with the lowest wind speeds (or more precisely, the lowest wind velocity vertical gradients). So, we expect that the best free seeing, isoplanatic angle, and coherence time, will all be found in this region, deteriorating with distance from there. This is in line with the findings of Hagelin \etal (2008), who predicted Dome A and South Pole to be comparable, with Dome F a little worse and Dome C significantly worse, because of high altitude winter winds.

\begin{figure}[hptb]
\epsscale{1.0}
\plotone{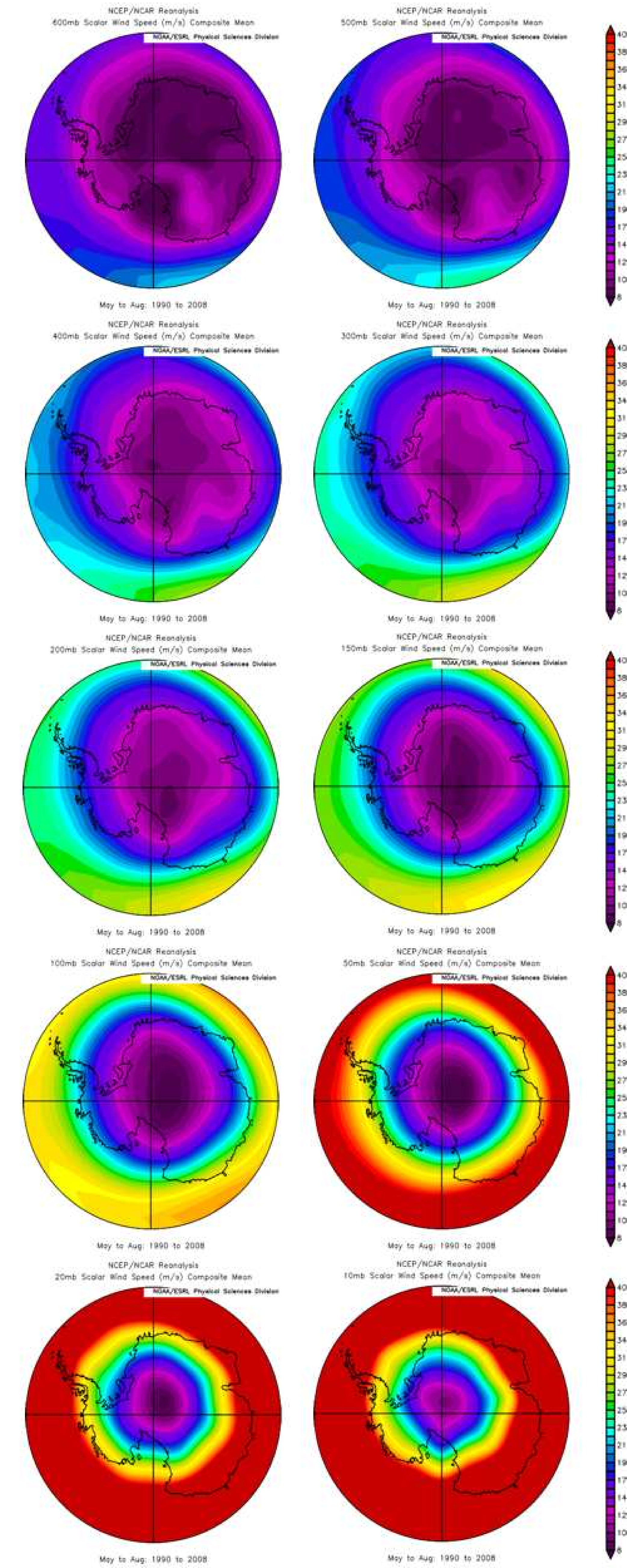}
\caption{Mean winter wind speeds at pressure heights 600, 500, 400, 300, 200, 150, 100, 50, 20, 10mb, for 1990-2008. Wind speed runs from 8km/s (purple) to 40km/s (red). \label{fig14}}
\end{figure}

We have tried to quantify the expected variation of the free seeing as follows. Firstly, following Masciadri and Jabouille (2001), SG06b, and Hagelin \etal (2008), we relate the refractive index variations $C_n^2$ to wind speed and potential temperature gradients, giving

\begin{equation}
C_n^2 = 3.62\times10^{-5} \left(\frac{1}{\sqrt{Pg}}\frac{1}{\theta}\frac{d \theta}{dh} \left| \frac{d\boldmath{V}}{dh}\right|\right)^{ 4/3}
\end{equation}

\noindent where $P$ is pressure, $g$ is gravitational acceleration, $\theta$ is potential temperature, $h$ is vertical height, $\boldmath{V}$ is velocity, and all terms are in SI units. Secondly, we make the very crude, but reasonable, assumption that the gradients $|d\boldmath{V}/dh|$  are proportional to the wind speeds taken from the NCAR data -- that is, that the atmosphere shows self-similar behaviour, with some fixed (though unknown) dependence between wind speed and its vertical gradients at any given pressure height. Thirdly, we use the winter NCAR/NCEP temperature and windspeed profiles over Dome C, together with the average winter $C_n^2$ profile of Trinquet \etal (2008, T08), as templates. We are then able to synthesise an average winter $C_n^2$ profile for every point in the NCAR maps, by scaling the T08 profile at each pressure height according to the local velocity and temperature profile. For elevations lower than Dome C, we have no $C_n^2$ profile, and we simply assume a fixed value of $C_n^2=1 \times 10^{-17}{\rm m}^{2/3}$ (m is metres). We then integrate up the resulting $C_n^2$ profile, to get maps for the free seeing $\epsilon_0$, the isoplanatic angle $\theta_0$, and also the coherence time $\tau_0$:

\begin{equation}
\epsilon_0 = 1.51 \times 10^{-15} \left(\int_{h_0} C_n^2 dh\right)^ {3/5}      [''] 
\end{equation}

\begin{equation}
\theta_0 =  9.48 \times 10^{15} \left(\int_{h_0} C_n^2 (h-h_0)^{5/3} dh \right)^{-3/5}   ['']
\end{equation}

\begin{equation}
\tau_0 =  4.60 \times 10^{17} \left( \int_{h_0} C_n^2 |\boldmath{V}|^{5/3} dh\right)^{-3/5}    [{\rm msec}]
\end{equation}

\noindent where $h_0$ is the surface height and all terms on the RHS are in SI units. The results are plotted in Figure 19. The variation due to differences in pressure and potential temperature profile are very small, it is wind speed that really matters. $\epsilon_0$ is dominated by contributions below about 11km (pressures above 200mb), while $\theta_0$ is dominated by contributions at  20-25km (pressures 20-40mb). So our maps are largely just a reflection of the average wind speeds at those heights. $\tau_0$ varies as a much stronger power of $|\boldmath{V}|$ than $\epsilon_0$ or $\theta_0$ ($\frac{9}{5}$ versus $\frac{4}{5}$), but is not dominated by any particular heights. The prediction for $\tau_0$ is also more robust, since there is a certain dependence on one power of wind speed. Table 6 shows the resulting values for the various sites. 

\begin{figure}[hptb]
\epsscale{1.0}
\plotone{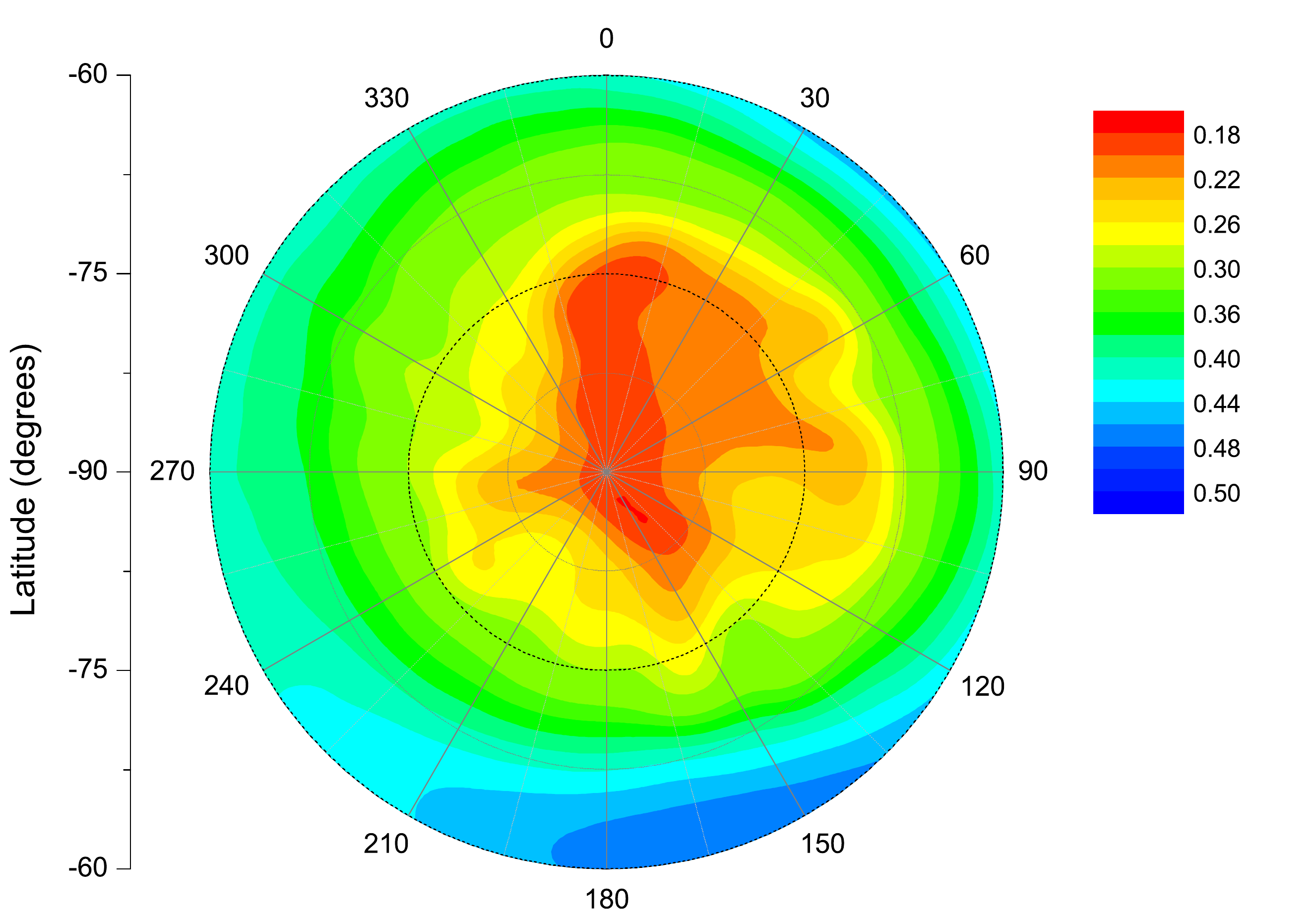}
\plotone{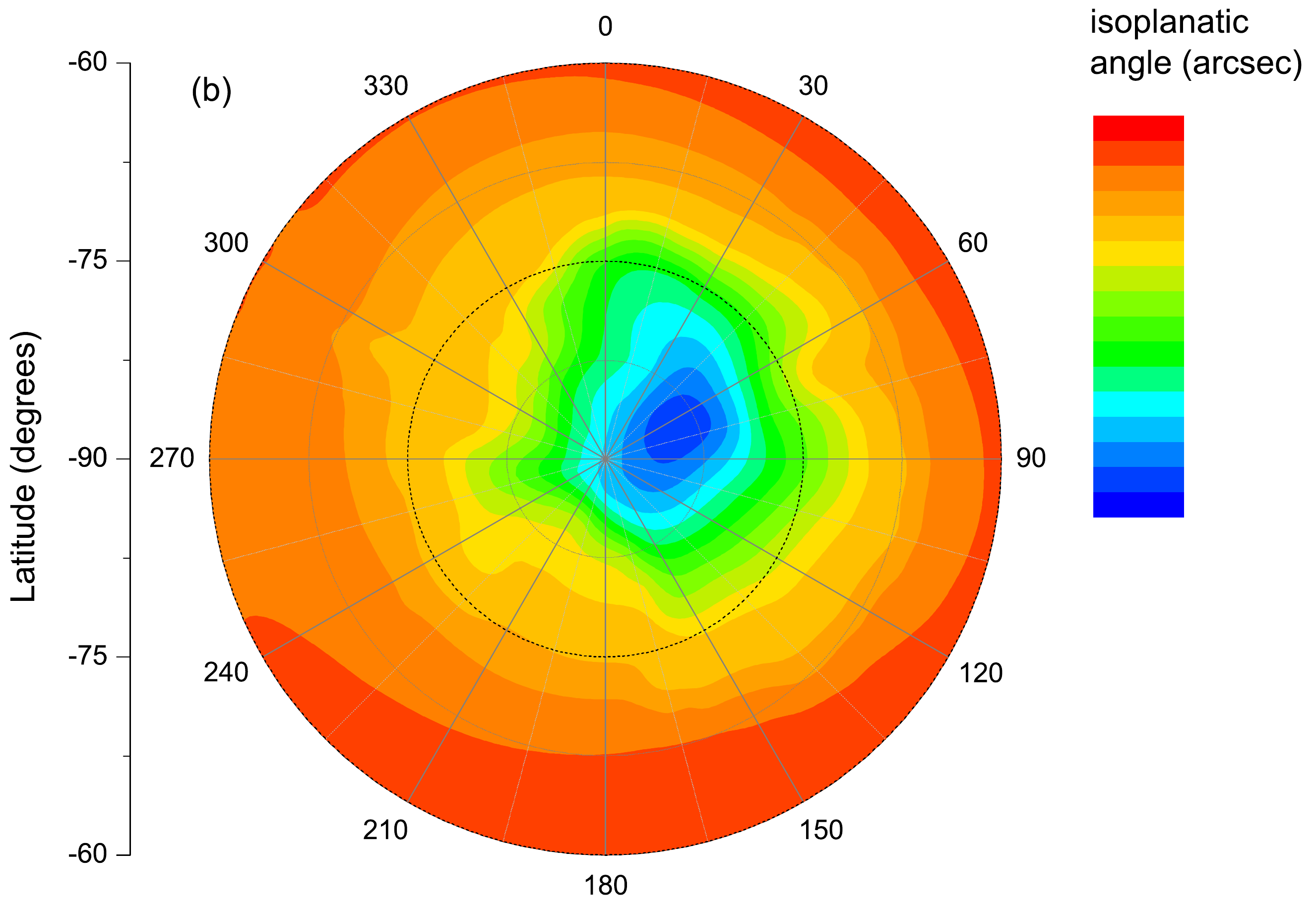}
\plotone{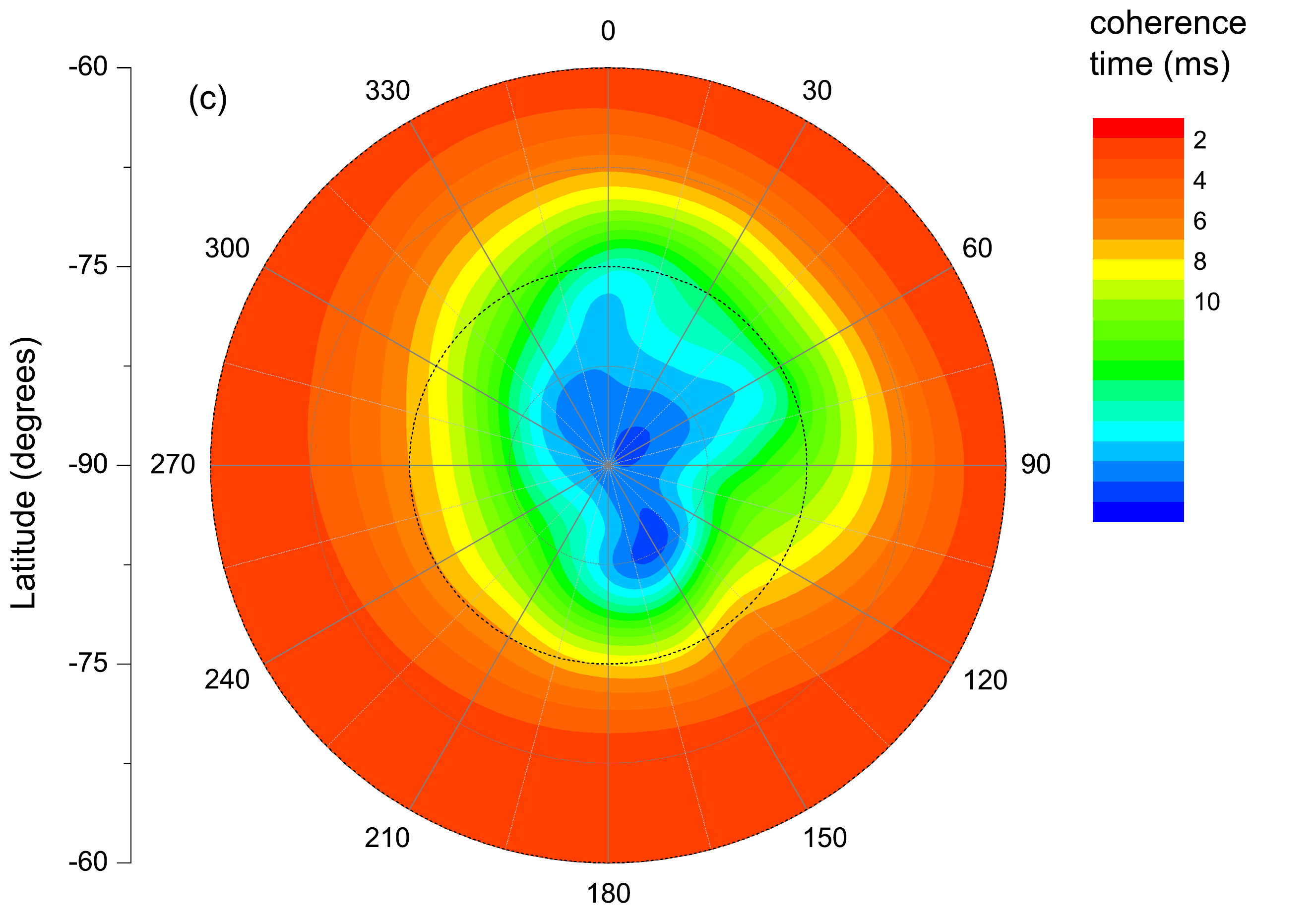}
\caption{(a, top) Predicted free seeing, (b, centre) isoplanatic angle, and (c, bottom) coherence time, all based on the model presented in the text. Orientation is 0E at the top, increasing clockwise, as per e.g. Figure 1. Latitudes 60\dg S, 67.5\dg S, 75\dg S and 82.5\dg S are marked. \label{fig14}}
\end{figure}

\begin{table}[hptb]
\begin{center}
\caption{Predicted free seeing $\epsilon_0$, isoplanatic angle $\theta_0$, coherence time  $\tau_0$, and the resulting `Coherence Volume' $\theta_0^2$ $\tau_0 / \epsilon_0^2$, for the various sites, using the model described in the text. \label{tbl-6}}

\begin{tabular}{lcccc}
\tableline\tableline
Site & $\epsilon_0$ (\as) & $\theta_0$ (\as) & $\tau_0$ (ms)& CV (s)\\ 
\tableline
South Pole & 0.186 & 5.43 & 18.8 & 16.0 \\ 
Dome A & 0.218 & 5.57 & 16.0 & 10.4 \\  
Ridge A & 0.208 & 6.29 & 17.6 & 16.1 \\ 
Dome C & 0.261 & 3.39 & 8.44 & 1.42 \\ 
Dome F & 0.209 & 5.17 & 15.4 & 9.42 \\  
Ridge B & 0.234 & 4.07 & 11.1 & 3.36 \\ 
\tableline
\end{tabular}
\end{center}
\end{table}

The model predicts that the best free seeing is at South Pole.  Comparing Dome A and Dome C, we predict that the free seeing is about 20\% better, the isoplanatic angle 50\% better, and for coherence time almost a factor of two better. Dome F is very nearly as good as Dome A, while Ridge A is significantly better than Dome A, and comparable with South Pole. Dome B is better than Dome C but much worse than Domes A or F. The model predicts very large variations in the utility of the sites for any sort of adaptive optics, which is given by the `Coherence Volume' $\theta_0^2$ $\tau_0 / \epsilon_0^2$ (e.g. Lloyd 2004). This is also shown in Table 6, and implies differences of an order of magnitude between Dome C and the best sites.

All the seeing predictions here look too optimistic, partly because the T08 profile gives better values than the DIMM measurements of Aristidi \etal (2009). Like Hagelin \etal (2008), we predict better free seeing at South Pole than at Dome C, in disagreement with the balloon data of Marks \etal (1999). We note that Lascaux \etal (in preparation) have undertaken simulations of individual $C_n^2$ profiles at Dome C, in impressive agreement with the T08 data. This offers the likelihood of much more sophisticated predictions for the comparative seeing of the various sites.

\section{Discussion}
The results on aurorae (based firmly on real data) and airglow (based on a model with only temperate validation), suggest that the optical sky brightness is higher in Antarctica than at temperate latitudes (except possibly briefly in the spring in the far red). The increase is not large enough to rule out e.g. interferometric or time-series observations, but makes Antarctica less attractive for sky-limited optical observations.

The cloud cover, surface temperatures, and high and low altitude winds all have minima offset from Dome A towards the South Pole, albeit by varying amounts. So it is natural to reconsider the topography in this region, to see if there is a better site for an astronomical observatory. Figure 20 shows the topography around Dome A, according to Liu \etal (2001). Dome A (80.37\dg S 77.53\dg E 4083m) is right on the northeast end of a very flat plateau. This is unfortunate, as there are better conditions at the other end of the plateau. There are two obvious other sites for an observatory: there is a secondary peak at (80.79\dg S 75.9\dg E 4080m), 55km away from Dome A but only 3m lower, and there is a perfectly flat spur which ends at (81.5\dg S 73.5\dg E 4053m), 144km from Dome A but only 30m lower. This latter site, which we call Ridge A, looks to offer significant advantages over Dome A in terms of weather, surface temperature, PWV, surface and high altitude wind speeds. 

\begin{figure}[hptb]
\epsscale{1.0}
\plotone{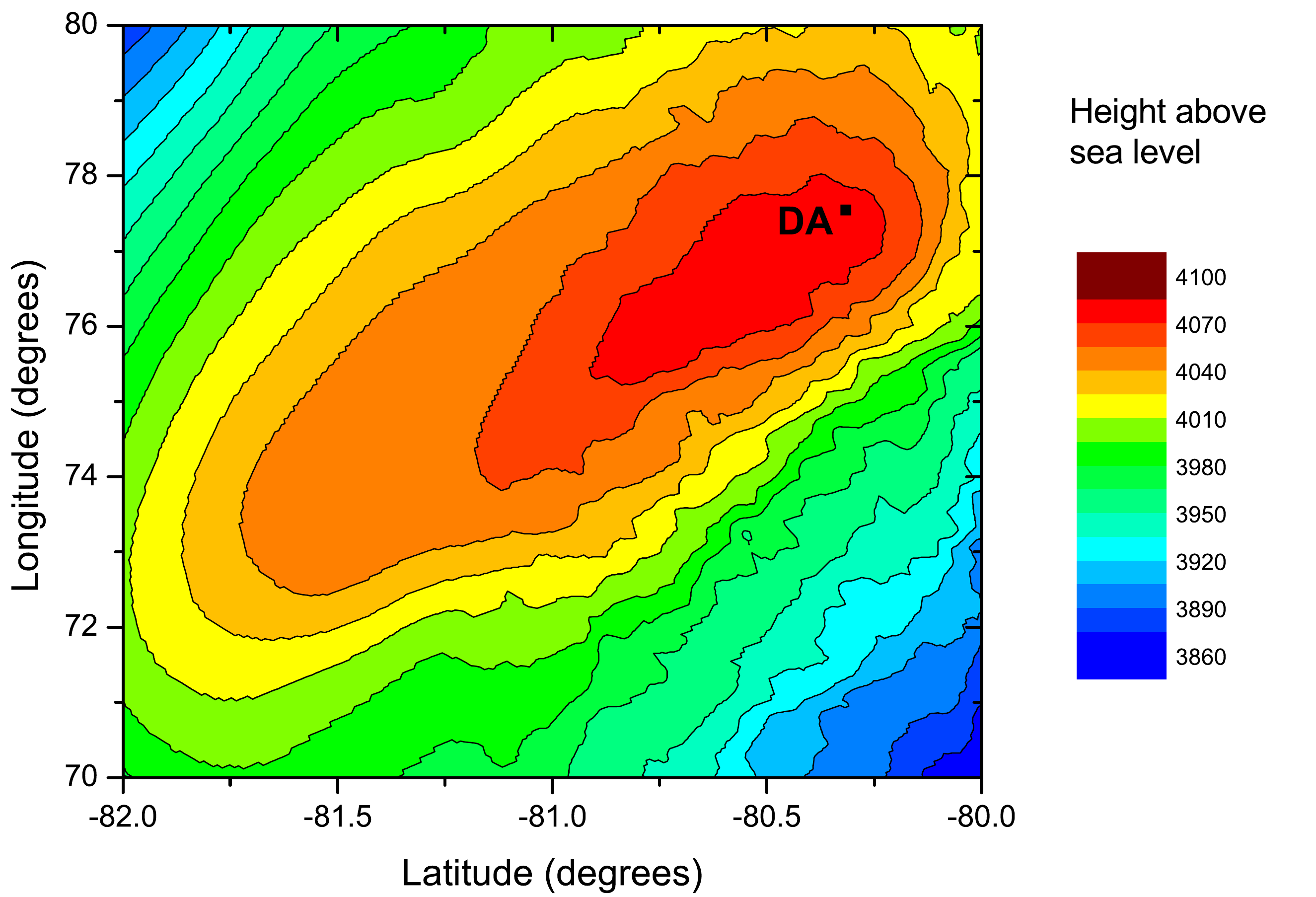}
\caption{ Surface contours in the vicinity of Dome A. Data from Liu \etal (2001), axes are in degrees. The lesser detail on the left hand side is an artifact of the available data resolution. \label{fig21}}
\end{figure}

Dome F emerges from this study as an excellent site, with the exception of auroral emission for optical work. The PWV is not quite as good as at Dome A, but the boundary layer, temperature, seeing, and weather characteristics are all comparable.

Ridge B (Figure 21), also contains potentially very good sites, if the problems of access and communications with Dome A prove intractable. Dome B at (79\dg 01\am S, 93\dg 37\am E, 3809m) has excellent boundary layer characteristics, is as high and cold as Dome F, with much lower auroral emissions, but with somewhat higher PWV and significantly worse free seeing. Positions further along the ridge (and so with better sky coverage and easier access) are compromised by the increasing mismatch between the physical peak, which runs almost due north, and the katabatic ridge running NNE. At the end of the katabatic ridge, at $\sim$76\dg S $\sim$97\dg E, the elevation is $\sim$3700m, $\sim$100m lower than the peak.

\begin{figure}[hptb]
\epsscale{1.0}
\plotone{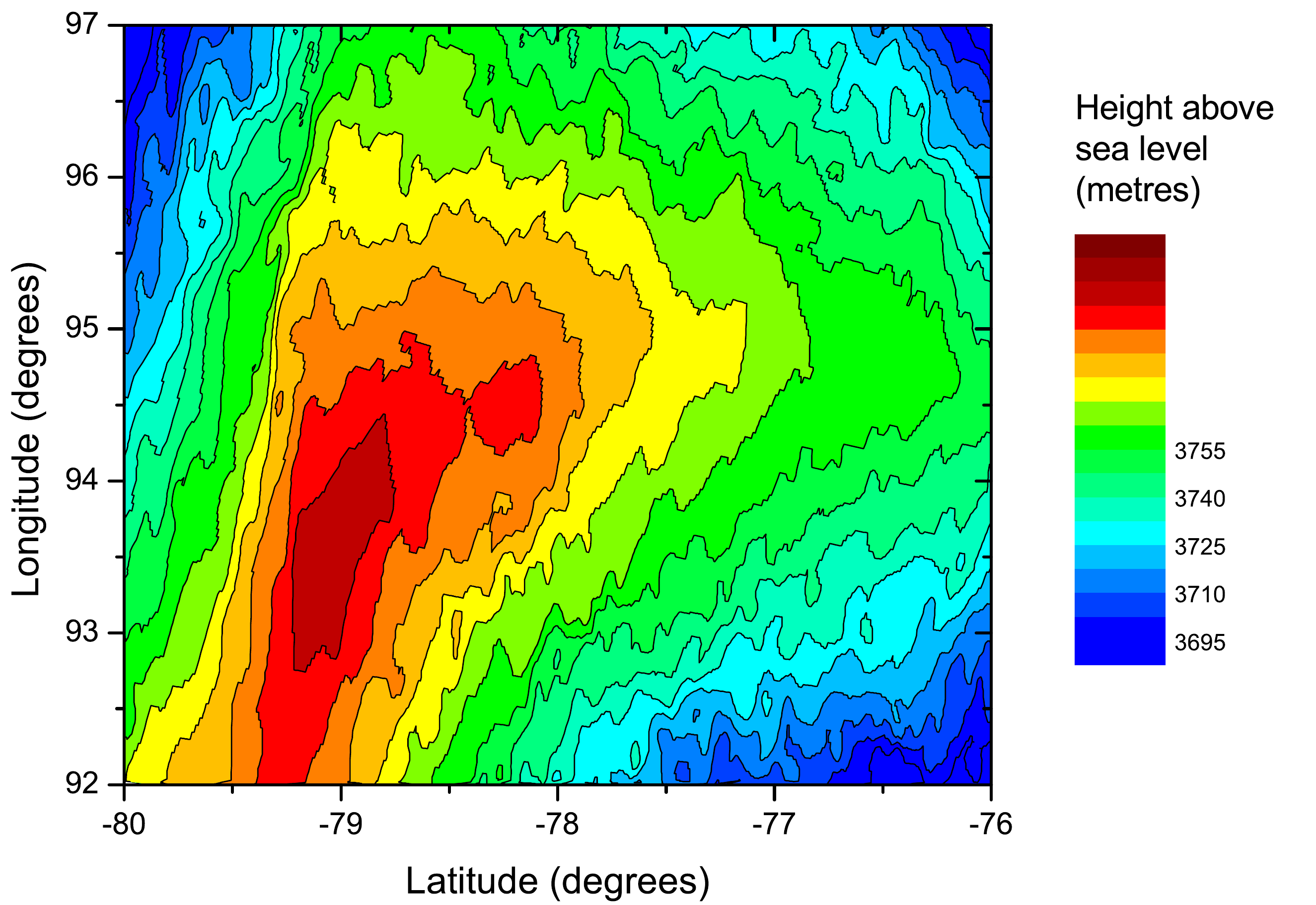}
\caption{ Surface contours in the vicinity of Ridge B. Data from Liu \etal (2001). \label{fig22}}
\end{figure}

Dome C scores very well in terms of surface temperature and weather, and is the only site able to use the predicted OH hole in the spring. The \Kdark background may be almost as good as other sites, though this is unmeasured. Seeing, thermal sky backgrounds and PWV are all significantly worse than Domes A or F.

South Pole appears to offer the least turbulence in the free atmosphere of all the sites, but is much poorer in almost all other respects.

\section{Conclusions}
\begin{enumerate}
\item{The lowest surface wind speeds and the thinnest boundary layer are found at Dome F, Ridge A, all along Ridge B, Dome C, and Dome A, in that order. The ridge on minimum wind speed is offset from the topographic ridge, in the direction of lower surface gradient.}
\item{The lowest surface temperatures are to be found along the line between Dome F -– Dome A -– Ridge B -- Dome C, with only a few degrees variation. The ridge of lowest temperature is again offset from the topographic ridgeline, but by a larger amount.}
\item{The lowest winter cloud cover closely tracks the lowest temperatures, with the same offset, and with only a few percent variation in the clear sky fraction. Dome C may have the best winter weather of all the sites.}
\item{The lowest wintertime free atmospheric wind speeds are found between Dome A, Dome F, and the South Pole, and increase with distance from there. On the assumption that wind speeds and seeing are closely correlated, this translates into significant differences in the free atmosphere seeing, the isoplanatic angle, and the coherence times. There is an order of magnitude variation in the predicted overall Coherence Volume between the sites.}
\item{The lowest wintertime atmospheric thermal emission, and the lowest precipitable water vapour, is likewise found between Dome A, Dome F, and South Pole. The differences between the sites in thermal IR sky background are factors of 1.5-3.}
\item{Domes A and C and Ridge B are all similar from an auroral point of view, with significant but not disastrous auroral contribution to the optical sky backgrounds. At Dome F, aurorae dominate the optical sky brightness.}
\item{Airglow from OI and OH is predicted to be higher everywhere on the Antarctic plateau than for temperate sites, limiting its attractiveness for sky-limited observations shortward of 2.2$\mu$m. However, in the spring, OH emission is predicted to collapse to levels $\sim5$ times lower than temperate sites. Dome C is the only site that can take advantage of this, but only for $\sim100$ hours/year.}
\item{Overall, Dome A is clearly the best of the existing sites, because of its excellent free atmospheric seeing, PWV, and thermal backgrounds from sky and telescope.}
\item{However, significantly better conditions are expected to be found $\sim$ 150km southwest of Dome A, at what we call Ridge A, at (81.5\dg S 73.5\dg E 4053m)}
\item{Dome F is a remarkably good site, comparable with Dome A, with the exception of PWV and auroral activity.}
\item{Dome B is also a very good site. The PWV is again not quite as good as Dome A, and the seeing is significantly worse. Dome F is marginally the better site for most purposes.}
\end{enumerate}

We summarise these conclusions as an entirely subjective, but hopefully useful, table of the merits of the various sites (Table 7).

\begin{table}[htpb]
\begin{center}
\caption{Scores for each site for the various criteria. \label{tbl-7}}
\begin{tabular}{|p{0.75in}|p{0.2in}|p{0.2in}|p{0.2in}|p{0.2in}|p{0.2in}|p{0.2in}|}
\tableline\tableline
 & SP & DA & RB & DC & DF & RA \\ 
\tableline%
Cloud & \cross & \tick & \tick\tick & \tick\tick\tick & \tick\tick & \tick\tick \\ 
Boundary Layer & \cross & \tick & \tick\tick & \tick & \tick\tick\tick & \tick\tick \\ 
Aurorae & \cross\cross &  \cross & \cross &  \cross &  \cross\cross\cross &  \cross \\ 
Free seeing & \tick\tick\tick & \tick\tick & \tick & \tick & \tick\tick & \tick\tick\tick \\
PWV / IR sky &  & \tick\tick & \tick\tick & \tick & \tick\tick & \tick\tick\tick \\  
Temperature & \cross & \tick\tick\tick & \tick\tick\tick & \tick\tick & \tick\tick & \tick\tick\tick \\  
\tableline
\end{tabular}
\end{center}
\end{table}

As a final comment, we note that the atmospheric properties of a site are only one set of characteristics that have to be considered before an observatory can be established.  To quote Vanden Bout (2002):

{\it ``Even so, there remain many other considerations that influence the selection of telescope sites.  These considerations can have an overwhelming influence.  They cannot be ignored if projects are to succeed.  The task of telescope site selection is to pick the site with the best atmosphere required for the science while satisfying the requirement that the project must be sold to those who will supply the funding.''}

\section{Acknowledgments} 
The weather satellite analyses were supported by the ICESat, CERES, and CALIPSO projects. We have made extensive use of NCAR/NCEP reanalysis data provided by the NOAA/OAR/ESRL PSD, Boulder, Colorado, USA, from their Web site at {\tt http://www.cdc.noaa.gov/}. Thanks to Yan Chen, Sharon Gibson and Petra Liverani for assistance with the graphics, and to Nicole van Lipzig for provision of a high quality version of Figure 4.

\end{document}